\documentclass[reprint,superscriptaddress,nofootinbib,amsmath,amssymb,aps,prl,onecolumn,longbibliography,notitlepage]{revtex4-1}

\usepackage[a4paper, margin=2.50cm]{geometry}

\usepackage{bbm}
\usepackage{graphicx}
\usepackage{fancyhdr, color}
\usepackage{hyperref} 
\usepackage[normalem]{ulem}
\usepackage{ragged2e}

\definecolor{DarkGreen}{rgb}{0,0.6,0}
\definecolor{amber}{rgb}{1,0.75,0}
\definecolor{brown}{rgb}{0.65, 0.16, 0.16}

\begin{document}

\fancyhead[C]{\sc Quantum thermodynamics of nanoscale thermoelectrics  and electronic devices}
\fancyhead[R]{}

\title{Quantum thermodynamics of nanoscale thermoelectrics and electronic devices}

\author{Robert S.~Whitney}
\email{robert.whitney@grenoble.cnrs.fr.} 
\affiliation{Laboratoire de Physique et Mod\'elisation des Milieux Condens\'es (UMR 5493), 
Universit\'e Grenoble Alpes and CNRS, Maison des Magist\`eres, BP 166, 38042 Grenoble, France.}

\author{Rafael S\'anchez}
\email{rafael.sanchez@uam.es.} 
\affiliation{Departamento de F\'isica Te\'orica de la Materia Condensada and Condensed Matter Physics Center (IFIMAC), Universidad Aut\'onoma de Madrid, 28049 Madrid, Spain.}

\author{Janine Splettstoesser}
\email{janines@chalmers.se.} 
\affiliation{Department of Microtechnology and Nanoscience (MC2), Chalmers University of Technology, S-412 96 G\"oteborg, Sweden}

\date{May 22, 2018}

\begin{abstract}
This chapter is intended as a short introduction to electron flow in nanostructures. Its aim is to provide a brief overview of this topic for people who are interested in the thermodynamics of  quantum systems, but know little about nanostructures.
We particularly emphasize devices that work in the steady-state, such as simple thermoelectrics, but also mention cyclically driven heat engines. We do not aim to be either complete or rigorous, but use a few pages to outline some of the main ideas in the topic.
\end{abstract}

\maketitle

\thispagestyle{fancy}

%% Introduction %%%%%%%%%%%%%%%%%%%%%%%%%%%%%%%%%%%%%%%%%%%%%%%%%%%%%%
\section{INTRODUCTION} \label{Sect:intro}
%%%%%%%%%%%%%%%%%%%%%%%%%%%%%%%%%%%%%%%%%%%%%%%%%%%%%%

{\it Thermoelectricity} is the name associated with any phenomena where a heat current induces an electric current, or vice versa.  It was much studied in the context of classical thermodynamics, with Onsager's Nobel-prize winning work on irreversible thermodynamic processes being first applied to the thermoelectric effect~\cite{Callen1985Sep}.
Thermoelectricity in \textit{nanostructures} was first observed experimentally in the early 1990's~\cite{Houten1992}, however it was little studied because the lack of good thermometry techniques at the nanoscale made quantitative experiments difficult. Now experimental progress in thermometry, see e.g.\
Refs.~\cite{Pekola2004Jun,Bourgeois2005Feb,Giazotto2006Mar,Reddy2007Mar,Heron2009Mar,Mavalankar2013Sep,Jezouin2013Nov,Kim2014Oct,Wang2018Jan}, has led to a renewed experimental interest, particularly in the use of nanoscale thermoelectrics to turn a heat flow into electrical power, or to turn electrical power into a heat flow from cold to hot (refrigeration).  This raises the question of developing a quantitative understanding of thermoelectricity in nanoscale structures, where quantum effects are important.

The objective of this chapter is to briefly explain how such thermoelectric effects occur, and discuss the quantum thermodynamics of these effects.  We do not intend our review to be complete. In particular, we restrict ourselves to nanostructures coupled to reservoirs of free electrons (i.e. metals or semiconductors), and  will not discuss the effect of superconductors at all. Further reading is proposed at the end of this introduction.

%%%%%%%%%%%%%%%%%
\begin{figure}[b]
\begin{center}
\includegraphics[width = 0.6\textwidth]{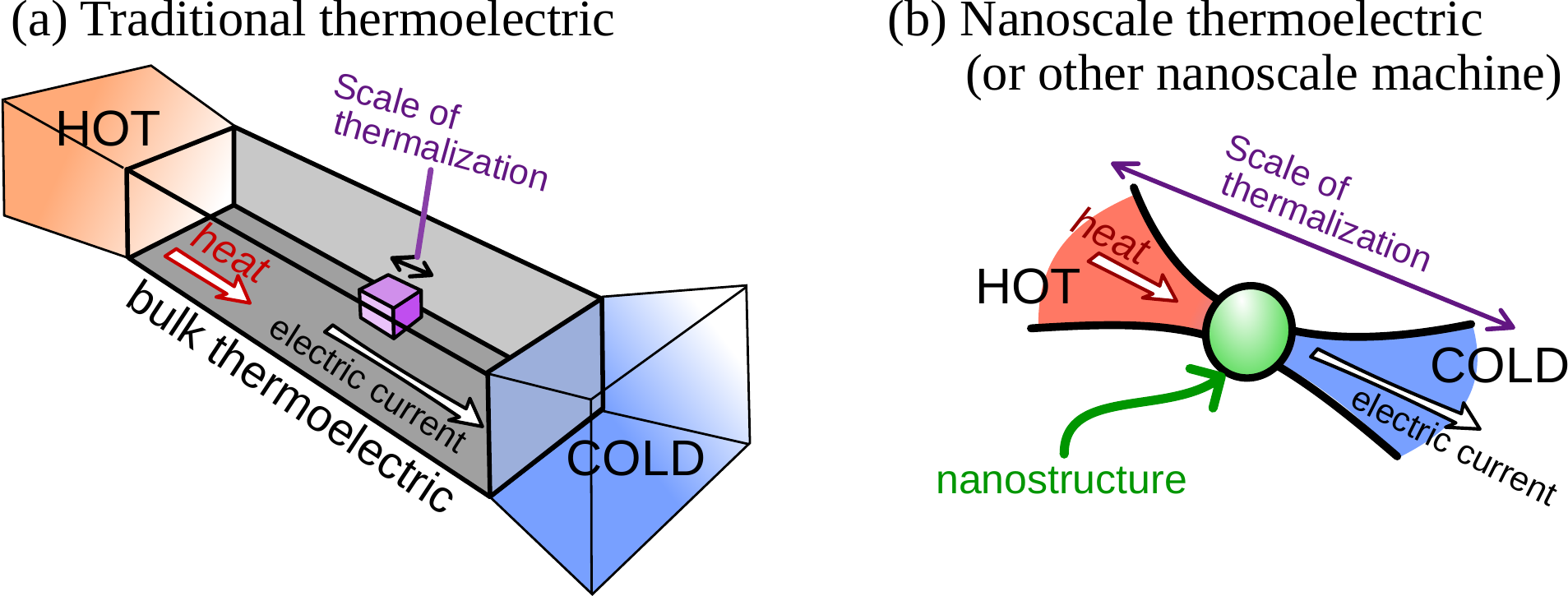}
\caption{\label{Fig:bulk-vs-nano} Figure adapted from Ref.~\cite{Benenti2017Jun}. In a traditional thermoelectric (a), the lengthscale on which the electrons relax to a local equilibrium is shorter than any lengthscale associated with the structure.  As a result, the electrons at each point in the structure can be treated as being in local thermal equilibrium, and have a local temperature which varies smoothly across the structure.  Then the system is well described by Boltzmann transport equations.  In contrast, in nanoscale thermoelectrics (b), or other nanoscale devices, the nanoscale structure is of similar size or smaller than the lengthscale on which electrons relax to a local equilibrium. Then the physics of the system can be much richer, exhibiting highly non-equilibrium effects.  The lack of local thermalization also means that the dynamics exhibit intrinsically quantum effects,
which would otherwise be destroyed by the decoherence that always accompanies thermalization.
}
\end{center}
\end{figure}
%%%%%%%%%%%%%%%%%%

%%%%%%%%%%%%%%%%%%%%%%%%%%%%%%%%%%%%%%%%%%%%%%%%%%%%%%
\subsection[]{Work generation in nanostructures}
%%%%%%%%%%%%%%%%%%%%%%%%%%%%%%%%%%%%%%%%%%%%%%%%%%%%%%

The main form of work produced by nanostructures is electrical; i.e.~they generate electrical power through a thermoelectric effect.
This work production involves moving electrons from a region of lower electro-chemical potential to a region of higher electro-chemical potential.  
This does work in exactly the same way as moving a mass up a hill (from lower gravitational potential to higher gravitational potential). It can be thought of as charging up a very large capacitor (turning heat into electrostatic work), charging up a battery (turning heat into chemical work) or using this potential difference to drive a motor (turning heat into mechanical work).  In all cases, if the nanostructure's thermoelectric effect moves $N$ electrons in a time $t$ from a region of electro-chemical potential $\mu_1$ to a region of electro-chemical potential $\mu_2$, then the electrical power it produces (the work it does per second) is $(\mu_2-\mu_1)N/t$.
In such a case, the electrical current is $I=eN/t$, where $e$ is the electronic charge, and one can define the voltage difference, $V$, via $eV=(\mu_2-\mu_1)$.  Then, we recover the familiar result that the electrical power produced is\footnote{The negative sign in $P$ is because we take the electric current $I$ to be positive when it flows 
from the reservoir at bias $V$ to the reservoir at zero bias.  Then $I$ has the same sign as $V$ when it flows ``downhill'' (from a region of higher bias to one of lower bias) turning electrical power into Joule heating.  This means that if the device is to {\it generate} power, then $I$ must have the opposite sign to $V$, so it is pushing electrical current ``uphill''.} $P=-VI$.

%%%%%%%%%%%%%%%%%%%%%%%%%%%%%%%%%%%%%%%%%%%%%%%%%%%%%%
\subsection[]{Traditional bulk semiconductor thermoelectrics}
%%%%%%%%%%%%%%%%%%%%%%%%%%%%%%%%%%%%%%%%%%%%%%%%%%%%%%

It is worth mentioning that bulk semiconductor thermoelectrics have been used to turn heat into electrical power for more than 40 years.
The most spectacular example being the Curiosity rover on Mars, which is a 900kg car-sized vehicle --- packed with scientific instruments --- powered  entirely by a thermoelectric generator (the heat source being the radioactive decay of a lump of plutonium-238).
Such thermoelectrics are used in space applications because their lack of moving parts makes them incredibly durable (the radioisotope thermoelectric generator on the space-probe Voyager 1 has been working for more than 40 years, despite now being outside the solar system). 
Thermoelectrics can equally be used for refrigeration via the effect known as Peltier cooling. 
However, to-date these power sources and refrigerators are too inefficient to be competitive for everyday earth-bound applications.  For example, the power source on the Curiosity rover has an efficiency of about 6\%, when the Carnot efficiency for the temperature difference in question would be of order 75\%.

We cannot pretend that nanoscale thermoelectrics are currently in a position to do better;
the efficiencies of the current experimental nanoscale thermoelectrics are tiny.  It is clear that a better understanding of nanoscale thermoelectric effects will lead to large increases in the efficiencies.  Whether they will get to the point of out doing bulk semiconductor thermoelectric remains to be seen.
A further expected advantage of nanoscale thermoelectric devices is however, that they provide ideas for on-chip refrigeration and waste-heat recovery for future nanoelectronic applications.

%%%%%%%%%%%%%%%%%%%%%%%%%%%%%%%%%%%%%%%%%%%%%%%%%%%%%%
\subsection[]{What is different at the nanoscale, and what is quantum?}
%%%%%%%%%%%%%%%%%%%%%%%%%%%%%%%%%%%%%%%%%%%%%%%%%%%%%%

If we compare nanostructures with traditional bulk solid-state devices, the main difference is that shown in
Fig.~\ref{Fig:bulk-vs-nano}. In nanostructures, all the interesting dynamics happens on a scale much smaller than the typical scale over-which electrons relax to a local thermal equilibrium.  Thus, the distribution of electrons in the nanostructure can be far from an equilibrium one.  This means that such systems cannot be described by the usual Boltzmann transport theory, and can exhibit the rich physics associated with highly non-equilibrium distributions.  At the same time, the fact that the nanostructure is of the size of the electron's wavelength, means that it often exhibits a quantization of energy levels similar to those in an atom; such structures are thus sometimes called {\it artificial atoms}. 
A major approach to design nanostructures is by patterning semiconductor heterostructures (and thereby partially depleting a 2-dimensional electron gas forming at one of the heterostructure's interfaces), see for example Ref.~\cite{Ihn2009Dec}. Another way of building nanostructures is to grow self-assembled quantum dots, or to place molecules with interesting discrete energy levels  between metallic contacts, particularly carbon nanotubes or large organic molecules.  

Many nanoscale structures are larger than a Fermi wavelength, and the electrons should be thought of as
free particles moving around in the nanostructure, bouncing off disorder or the nanostructure's boundaries, tunnelling through barriers, etc.  However, such a nanostructure is often smaller than
the lengthscale over which electrons decohere; in other words, the electrons can pass through the nanostructure without losing the phase of their quantum wavefunction. 
This means their dynamics can exhibit interference effects analogous to those in optics (Young's interference, Fabry-Perot, Mach-Zehnder, etc.) as well as the famous Aharonov-Bohm effect, see e.g. Ref.~\cite{Nazarov2009May}.  Such effects cannot be described by the classical transport theories (such as Boltzmann transport theory) used for bulk systems.

In principle, the long coherence times of the electron wavefunctions also mean that entanglement generated between electrons (through their interactions with each other inside the nanostructure) will survive long enough to have effects on the device's operation. However, clear cut observations of entanglement has proved more difficult than observing interference effects.  Some steps towards devices that involve entanglement can be seen in
Refs.~\cite{Karimi2017Aug,Brunner2012May,Hofer2016Dec}.

%%%%%%%%%%%%%%%%%%%%%%%%%%%%%%%%%%%%%%%%%%%%%%%%%%%%%%
\subsection[]{What else can we learn from thermoelectrics?}
%%%%%%%%%%%%%%%%%%%%%%%%%%%%%%%%%%%%%%%%%%%%%%%%%%%%%%

As with other systems studied in quantum thermodynamics, to find ways of generating a large power with high efficiency is not the only question.  Indeed, much of the field is focused on the more fundamental issue of understanding the physics of thermoelectric nanostructures.  In particular,
we know that the thermoelectric response of a system gives us access to different information from a measurement of its charge conductance.  
For example, in systems described by Landauer-B\"uttiker scattering theory\footnote{In Landauer-B\"uttiker scattering theory, the mesoscopic electronic device is viewed as a scatterer onto which electronic wavefunctions, incoming from the leads, impinge and are transmitted or reflected. This well-known powerful approach is particularly useful for systems with weak Coulomb interaction and underlies the reasoning of various sections of this chapter. Details can be found in various text books, see for example~\cite{Moskalets2011Sep,Datta1997May,Heikkila2013}.} the thermoelectric response
depends on the difference between the dynamics of electrons above the Fermi surface from the dynamics
of electrons below the Fermi surface \cite{Sivan1986Jan, Butcher1990} (i.e., on broken electron-hole symmetry), when the electrical conductance only depends on the sum of the two.
Thus, thermoelectric effects can be used as novel probes of the physics happening within nanostructures.
However, to make such a probe quantitative, one needs good quantitative models of the thermoelectric responses of all kinds of nanostructures, regardless of whether those responses are large or small.

Examples of uses of thermoelectric effects to study the physics of nanostructures include:
the detection of interaction between different channels in the quantum Hall regime~\cite{Ronetti2016Apr},
the presence of odd superconducting states~\cite{Hwang2017Dec},
the scale on which energy relaxation occurs~\cite{Granger2009Feb,Nam2013May}, signatures of exotic or topological states~\cite{Sothmann2016Aug,Erlingsson2017Jul}, the existence of neutral modes in fractional quantum Hall states~\cite{Gross2012May,Altimiras2012Jul}, and anyonic currents~\cite{Banerjee2017Apr}.

%%%%%%%%%%%%%%%%%%%%%%%%%%%%%%%%%%%%%%%%%%%%%%%%%%%%%%
\subsection[]{Further reading}
%%%%%%%%%%%%%%%%%%%%%%%%%%%%%%%%%%%%%%%%%%%%%%%%%%%%%%

For readers interested in more details on thermoelectric effects in nanostructures, we can suggest
the review in Ref.~\cite{Benenti2017Jun}.  
The thermoelectricity of structures containing superconductors is discussed in Refs.~\cite{Giazotto2006Mar,Muhonen2012Mar}, while the quantum thermodynamics of other superconducting circuits are discussed elsewhere in this book. 
Reviews of quantum dots and their potential for thermoelectric effects are given in Refs.~\cite{Sothmann2015,Haupt2013Jun}.  
Thermoelectric effects in atomic and molecular junctions are reviewed in 
Refs.~\cite{Dubi2011Mar,Ratner2013Jun,Bergfield2013Nov}.  Ref.~\cite{Pichard2016Dec} gives a taste of a variety of topics related to mesoscopic thermoelectrics. 
Ref.~\cite{Ventra2008Aug} is a textbook that overviews a number of theories of transport in nano-systems; those  discussed here and others. 

It is also worthwhile having a basic overview of  thermoelectricity in bulk systems, such as provided in the textbooks~\cite{Ioffe1958Dec,Goldsmid2009Oct,Rowe1995Jul}.
Useful reviews on bulk thermoelectrics include Refs.~\cite{DiSalvo1999Jul,shakouri2009,Shakouri2011Jul,Pop2006Aug},
with perspectives for nano-structured bulk materials in Refs.~\cite{Koumoto2013Aug,Macia2015May}.
Thermal transport at the nanoscale is reviewed 
in Refs.~\cite{Cahill2003Jan,Cahill2014Mar}.

%%%%%%%%%%%%%%%%%%%%%%%%%%%%%%%%%%%%%%%%%%%%%%%%%%%%%%
\section{Energy selection for heat to work conversion}
%%%%%%%%%%%%%%%%%%%%%%%%%%%%%%%%%%%%%%%%%%%%%%%%%%%%%%

%%%%%%%%%%%%%%%%%
\begin{figure}[t]
\begin{center}\includegraphics[width = 0.9\textwidth]{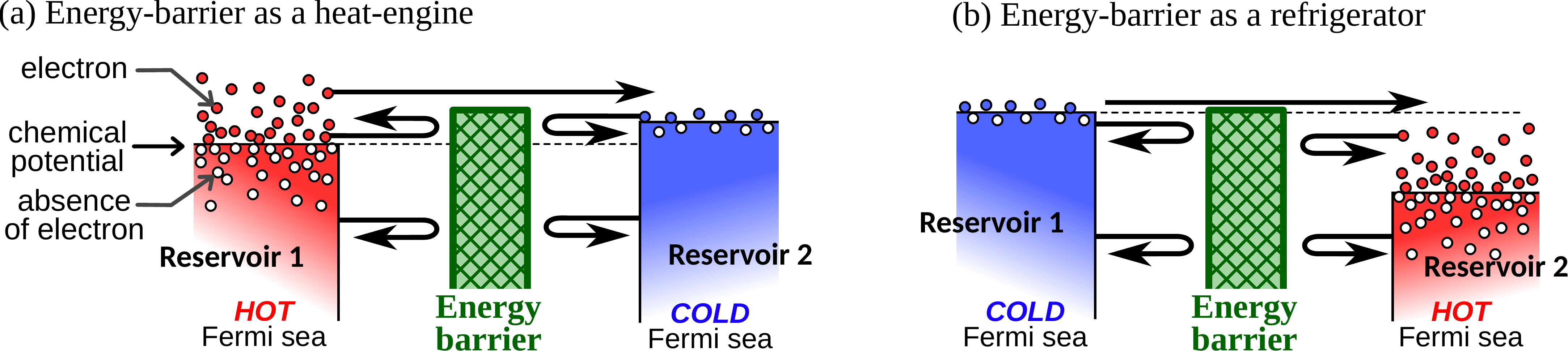}
\caption{\label{fig:energy-filter} A sketch, adapted from Ref.~\cite{Benenti2017Jun}, which shows energy selection via an energy barrier, which induces a thermoelectric effect.  At zero temperature, electronic states in the reservoirs would be filled up to the Fermi energy, which is indicated by the upper limits of the red and blue regions. Non-vanishing temperatures lead to thermal electron-hole excitations modifying the occupation in the reservoirs. This simple set-up is used as a heat engine in (a); it generates power because the temperature difference means that electrons flow from a hot region of lower electrochemical potential to a cold region of higher electrochemical potential. It is used as a refrigerator in (b);  it uses the potential bias to ensure that electrons above the Fermi sea can flow out of the cold reservoir, cooling it further.}
\end{center}
\end{figure}
%%%%%%%%%%%%%%%%%%

The simplest solid-state device to convert heat into work (or to use work to move heat from cold to hot)
is one that selects energies, by only allowing certain energies to flow through it (thereby acting as an energy filter). 
An energy-barrier is the simplest such energy selective system, an example is sketched in Fig.~\ref{fig:energy-filter}.  All particles with energies above the top of the barrier can flow freely between hot and cold, while all those with energies below cannot, so the barrier is a {\it high-pass} energy filter.\footnote{We know from the textbook problem of a quantum particle hitting a barrier, that the transmission probability will be a smooth function of energy (going from  zero at low energies to one at high energies), because the solution of the wave equation allows for tunnelling through the barrier.}
This energy filter acts as a {\it thermoelectric}, because a heat current induces a charge current  and vice-versa.

While bulk semiconductor thermoelectrics are not the subject here, we do note that they function in a similar manner.
Their energy selection induced by their band-structure, with charge carrier flowing at energies in a band, and not flowing at energies in a band gap. 
The physics is a little different because thermalization occurs inside the structure (with the local temperature dropping uniformly from hot to cold), but is well captured by Boltzmann transport theory \cite{Goldsmid2009Oct,Mahan1996Jul}.
 
A crucial question when exploiting the exotic features of nanostructures for thermoelectrics,
is how the well-known laws of thermodynamics apply.  On a practical level, it  is generally assumed that physical systems will not  violate these laws, which raises the question of verifying that theoretical models do not violate them either.  On a more philosophical level, it is believed that  the laws of thermodynamics (particularly the second law) are an emergent property of more fundamental laws of physics. Thus, one can ask if they can be derived from
a suitable quantum theory.
This can be done relatively easily in the context of quantum thermoelectrics with weak enough electron-electron interactions that such interactions can be ignored (both in the reservoirs and in the nanostructure), see \textsf{Box 1}.  Then, the steady-state flow through the nanostructure can be modelled with Landauer-B\"uttiker scattering theory.  It is not hard to show, within this theory, that the dynamics will not violate the laws of thermodynamics.  The detailed proofs are given elsewhere, such as in Ref.~\cite{Benenti2017Jun}
and we just briefly outline them 
here, 
\begin{itemize}
\item {\bf First law of thermodynamics.} An electron with energy $E$ passing from reservoir 1 to reservoir 2 through the nanostructure conserves its energy.  In contrast, this process generates a work of $\mu_2-\mu_1$, where $\mu_1$ and $\mu_2$ are the electro-chemical potentials of the two reservoirs.  However, this process removes a heat equal to $E-\mu_1$ from reservoir 1 and adds a heat $E-\mu_2$ to reservoir 2.  
Thus, the sum of the heat and work generated by the process is zero,
and we have the first law of thermodynamics.
\item{\bf Second law of thermodynamics.}  Following the Clausius definition, the entropy change in reservoir $j$ equals the
change in that reservoir's heat multiplied by its inverse temperature $\beta_j$ (for simplicity we measure entropy in units of $k_{\rm B}$).   In Landauer scattering theory, the flow of electrons at energy $E$ from reservoir 1 to reservoir 2 is ${\cal T}(E) \big[(f(x_1)-f(x_2)\big]$, where ${\cal T}(E)$ is the transmission probability at energy $E$, and $f(x_j)$ is the Fermi distribution function for electrons in the reservoir $j$ with $x_j=\beta_j(E-\mu_j)$.   
The total entropy change (change in reservoir 1 and 2) due to electron flow at energy $E$ is 
thus proportional to $(x_2-x_1)\big[f(x_1)-f(x_2)\big]$.  As $f(x)$ is a monotonically decaying function of $x$, this is never negative, and the second law is guaranteed.
\end{itemize}

Given the proof of the second law, it is not hard to guess that the only way to make the system reversible in the thermodynamic sense (i.e. to ensure it generates no entropy) is to only allow transmission between reservoir 1 and 2 at the one energy
where the Fermi distribution functions in reservoir 1 and 2 are equal. 
We call this reversibility energy $E^\rightleftharpoons$ and it equals $\mu_1 + \beta_2 (\mu_2 -\mu_1)/(\beta_2-\beta_1)$.  The fact that the occupation function for reservoir states is the same in the two reservoirs at this energy, means that the same number of particles flow from 1 to 2 and from 2 to 1,
and so there is no heat current or electrical current.  However, if one allows particles to flow in a vanishingly narrow energy window just above $E^\rightleftharpoons$ then there will be an infinitesimal current, which converts heat to work with infinitesimal entropy production, so the efficiency of the conversion is 
arbitrarily close to Carnot efficiency~\cite{Humphrey2002Aug,Nakpathomkun2010Dec}.  
Such an ideal energy-selection can be constructed with a quantum dot or molecule with a single-level at an energy 
tuned to be at $E^\rightleftharpoons$. Systems similar to this have been made experimentally, 
for a recent review of quantum dots see \cite{Svilans2016Dec}, and for a discussion of molecules see \cite{Reddy2007Mar,Kim2014Oct,Cui2017Dec}. 
There are various practical reasons why these systems do not reach Carnot efficiency, the most difficult one to resolve is the flow of heat from hot to cold through phonons and photons, which parasite the efficiency. Note that designs to improved bulk thermoelectrics (for example by nanostructuring them), by giving them peaked spectra have been proposed and heavily exploited~\cite{Mahan1996Jul,Hicks1993May,Hicks1993Jun,Humphrey2005Mar,Heremans2013Jun}.

%===================================================
\begin{table}[b]
\fbox{\hskip 3mm
\begin{minipage}{0.9\textwidth}
\sffamily
\vskip 2mm
\textbf{BOX 1: INTERACTIONS BETWEEN ELECTRONS IN NANOSTRUCTURES}
\vskip 2mm
\justify
Interactions between electrons in nanostructures are counter-intuitive; electrons interact strongly with each other when their density is \textit{low}, but interact weakly when their density is \textit{high!}  
This is a consequence of them being fermions, and is formally described by Landau-Fermi liquid theory.  However, we can get a hint of why this is from a simple argument comparing the kinetic energy to the Coulomb interaction energy.
If we consider $N$ free electrons in a $d$-dimensional box of size $L$, they form a Fermi surface whose Fermi momentum $p_F$ is given by the equation $N \sim (p_FL/h)^d$ (we drop all factors of order one).
Thus, the kinetic energy per electron at the Fermi surface goes like  $N^{2/d}$.
In contrast, the Coulomb interaction energy goes like $1/r$, and the typical distance between particles $r$ goes like $N^{-1/d}$,  hence the Coulomb energy per electron goes like $N^{1/d}$.
Thus, for large $N$, the kinetic energy will dominate over the Coulomb interaction energy, and one can guess
that interaction effects will be weak.  Landau-Fermi liquid theory gives a quantitative theory of this, and explains that screening effects enhance this suppression of the Coulomb interaction energy at high-density.

In the context of nanostructures, this means that we can often treat the reservoirs (which typically have a high density of electrons) as containing non-interacting electrons.  However, the nanostructure itself can have a low density of electrons confined in a manner that reduces their screening. 
Thus, Coulomb interaction effects between the electrons in the nanostructure can be significant, and can lead to various effects, such as \textit{Coulomb blockade} or \textit{Luttinger liquid} physics.  Alternatively, the nanostructure could consist of multiple  regions with high electron densities,  separated by barriers (where the electron density is low or even zero) which block the free flow of electrons.  Then electrons are approximately non-interacting within each region, but there are strong interactions between the electrons in the different regions, just like the well-known interaction between capacitor plates.

The exact value of the Coulomb interaction in a given nanostructure can often be hard to predict, because it depends a lot on the screening due to the electron gases in the vicinity of the nanostructures (for example the electron gases in the reservoirs).  It is thus typically quantified in terms of phenomenological capacitances, which must be measured in the nanostructure in question.  

$\ $
\end{minipage}\hskip 3mm}
\end{table}
%======================

%%%%%%%%%%%%%%%%%%%%%%%%%%%%%%%%%%%%%%%%%%%%%%%%%%%%%%
\subsection[]{Quantum bounds on efficiency at finite power output}
%%%%%%%%%%%%%%%%%%%%%%%%%%%%%%%%%%%%%%%%%%%%%%%%%%%%%%

The fact that electrons are quantum objects places bounds on
a thermoelectric's power output and efficiency \cite{Whitney2014Apr,Whitney2015Mar,Whitney2016May}, 
which do not appear in classical thermodynamics. These {\it quantum bounds} come from the wave-like nature of the electrons. 
The first observation of this type was due Pendry \cite{Pendry1983}, derived from the Landauer scattering theory by observing that heat flow is maximized by having all electrons in each mode being transmitted at all energies.
Then the upper bound on the heat current out of a reservoir at temperature $T$  \cite{Pendry1983}, 
\begin{eqnarray}
\label{maxJ}
J_{\rm m} = {\pi^2 \over 6 h} \ N \ \big(k_{\rm B} T\big)^2.
\end{eqnarray}
Here, $N$ is the number of modes in the cross-section through which that heat current is flowing,
which is given by the cross-section measured in units of the Fermi wavelength.
This is much like the Stefan-Boltzmann law for heat carried by the photons emitted from a black-body.
This upper bound is due to the wave-like nature of the electrons, because if one takes the electron wavelength to zero, 
the number of modes $N$ in a given cross-section diverges, and the above limit becomes irrelevant.
Pendry's work \cite{Pendry1983} also used this to point out that the rate of reduction of entropy in a reservoir, $-({\rm d} {\cal S}/{\rm d}t)$, cannot exceed 
$J_{\rm m}/T$, and made the connection to the amount of information that can flow in an $N$ mode channel, but we do not discuss that further here.

As there is also an upper bound on a heat-engine's efficiency (given by Carnot efficiency), the above upper-bound on heat current directly implies an upper-bound on the heat engine's power output.
A more detailed look at the Landauer scattering theory makes one realize there is a contradiction between 
maximizing the heat flow (which occurs when electrons flow from hot to cold at all energies) and maximizing 
the efficiency of a heat-engine (which requires only letting electrons flow at one specific energy).  
Refs.~\cite{Whitney2014Apr,Whitney2015Mar,Whitney2016May} found the compromise between these two limits which maximizes the power output of a thermoelectric. That maximum, which depends only on temperature and universal constants, is
\begin{eqnarray}
P^{\rm max}_{\rm output} = {A_0 \pi^2 \over h} \ N \ k_{\rm B}^2 \big( T_{\rm hot}^2-T_{\rm cold}^2\big)
\end{eqnarray}
where $A_0\simeq 0.032$.  This was recently probed experimentally in a InAs nanowire \cite{Chen2018Apr}, however to get a feeling of its implications, it is easier to apply it to a more macroscopic situation.  We can consider the case of recovering energy from the waste heat in car exhaust ($T_{\rm hot}\sim$ 700\,K and $T_{\rm cold}\sim$ 300\,K) and a Fermi wavelength of order 10\,nm (such as in a typical semiconductor).  This equation means that
a millimetre-square cross-section (which thus carries  $N \sim 10^{10}$ modes) cannot generate more than
about 300\,W of power. Remarkably, it is quantum mechanics
which gives this bound (since it would be infinite if one sets the electron wavelength to zero), even though the cross-section and power outputs 
in question are macroscopic.   A power output of 300\,W per mm$^2$ seems fairly large, until one realizes that a filament lamp has a filament whose cross-section is ten thousand times smaller at $10^{-4}\,$mm$^2$ yet it can easily carry 100\,W.

Refs.~\cite{Whitney2014Apr,Whitney2015Mar,Whitney2016May} show that there is a more stringent bound on efficiency than Carnot's bound at any finite power output. A device can only get to an efficiency as large as Carnot efficiency if the power output is much less than $P^{\rm max}_{\rm output}$.
The efficiency bound deviates from Carnot efficiency by a factor which goes like
$\big(P_{\rm output}\big/P_{\rm output}^{\rm max}\big)^{1/2}$  for $P_{\rm output}\ll P_{\rm output}^{\rm max}$, so Carnot efficiency is only strictly achievable when $P_{\rm output}\to 0$.
A more recent work \cite{Brandner2015Jan}, considered the effect of a  large amount of relaxation inside the scatterer, and suggested that the deviation from Carnot efficiency might be smaller (going like $P_{\rm output}\big/P_{\rm output}^{\rm max}$ for $P_{\rm output}\ll P_{\rm output}^{\rm max}$). 
However, this still means that Carnot efficiency is only strictly achievable when $P_{\rm output}\to 0$. This observation is due to the quantum nature of electrons, if one took the classical limit by taking their wavelength to zero, $P^{\rm max}_{\rm output}$ would go to infinity, and Carnot efficiency would be achievable at
any $P_{\rm output}$.
It is not yet clear if these bounds apply beyond scattering theory, such as when there are significant Coulomb blockade effects.

%%%%%%%%%%%%%%%%%%%%%%%%%%%%%%%%%%%%%%%%%%%%%%%%%%%%%%
\subsection[]{Linear response, the Seebeck coefficient and the figure of merit}
\label{sec:linear}
%%%%%%%%%%%%%%%%%%%%%%%%%%%%%%%%%%%%%%%%%%%%%%%%%%%%%%

In the context of bulk thermoelectrics, it is common to talk about the Seebeck coefficient $S$, 
and the dimensionless figure of merit, $ZT$.
The Seebeck coefficient is a measure of the strength of the thermoelectric effect.\footnote{The Seebeck coefficient $S$ is often called the thermopower, even though it does not have either the meaning or the units of power.}
The dimensionless figure of merit  is a quantity that identifies the maximal thermodynamic efficiency of the thermoelectric.  Both these quantities only have a sense in the so-called {\it linear response regime}, which occurs  when the 
bias, $V$, and temperature difference $\Delta T$, across the sample are small enough, that the electrical current is linear in $V$ and $\Delta T$.  
This is typically the case when the temperature drop on the scale of the
distance over which thermalization occurs is small compared with the temperature.  In bulk thermoelectrics, this is almost always the case, see Fig.~\ref{Fig:bulk-vs-nano}a, and so most of the literature discusses how to optimize $S$ and $ZT$.   In contrast, nanostructures leave the standard linear response regime as soon as the ratios $\Delta T/T$  or $eV\big/ k_{\rm B}T$ cease to be small.\footnote{In some special cases, one can construct a more exotic linear response theory when $k_{\rm B}\Delta T$  or $eV$ are small compared to an energy scale which is not temperature.}
Let us again refer to the commonly cited example of an application to generate electricity from the heat in the exhaust gases of a car at 600-700K
when the surrounding temperature is at 270-300K; clearly a nanostructure that could do this
would be operating far from the linear response regime, since $\Delta T/T$ is obviously not small.    
Thus, this review mainly discusses modelling that goes beyond linear-response.

However, in many cases the first attempts to model a nanostructure's thermoelectric effect are in the linear-response regime, because it is simpler.  So everyone interested in thermoelectrics should have a basic understanding of this regime, and the specific results one can derive there.  This section briefly summarizes the most important ones, such as explaining the Seebeck coefficient $S$, 
and the dimensionless figure of merit, $ZT$.

In the thermodynamics of systems in the linear response regime,  one can write the 
electrical current $I$ and heat current $J$  in terms of the thermodynamic forces, $V/T$ and $\Delta T\big/T^2$, as follows
\begin{eqnarray}
\left(\begin{array}{c} I \\ J\end{array}\right) &=& 
\left(\begin{array}{cc} L_{11} & L_{12} \\  L_{21} & L_{22} \end{array}\right) \left(\begin{array}{c} V/T \\  \Delta T\big/T^2 \end{array}\right) 
\end{eqnarray}
where the matrix is the Onsager matrix \cite{Callen1985Sep}.
In the context of thermoelectric systems  one tends to write this matrix in terms of parameters commonly measured in experiments.
Then it takes the form of the pair of coupled equations
\begin{eqnarray}
I \ &=&\ G\,V \ +\ GS \, \Delta T,
\label{Eq:I-linear} \\
J \ &=&\ G\Pi\,  V \ +\ (K+ GS\Pi)\, \Delta T .
\label{Eq:J-linear}
\end{eqnarray}
These two equations contain four parameters with the following experimental meanings.
The electrical conductance, $G$, is defined as the ratio $I/V$ when there is no temperature difference, $\Delta T=0$.
The Seebeck coefficient, $S$, is a measure of the voltage generated across a thermoelectric by a temperature difference, when that thermoelectric is not connected to an electrical circuit, so $V$ takes the value that ensures that $I=0$. Hence, $S$ defines the thermovoltage $V_{\rm s}=-S\Delta T$\footnote{There is ambiguity in the literature about the sign of $S$. One can choose either sign, as long as one is consistent.} at which no charge current flows ($V_{\rm s}$ is also known as the stopping voltage).
The Peltier coefficient, $\Pi$, is a measure of the directional heat flow induced by an electric current, when $\Delta T=0$.  Hence, $\Pi$ equals the ratio $J/I$ when $\Delta T=0$.    Onsager showed that $\Pi=T S$ in the absence of an external magnetic field (or more generally, for systems with time-reversal symmetry) \cite{Callen1985Sep}.  
Finally, $K$ is the thermal conductance, defined as the Fourier heat flow divided by $\Delta T$, where the Fourier heat flow is the heat flow when the  thermoelectric is not connected to an electrical circuit,
so $V$ takes the value that ensures that $I=0$; hence $K=J/\Delta T$ under the assumption that $V=V_{\rm s}$.

If one can find $G$, $S$, $\Pi$ and $K$ for a given nanostructure, either through experimental measurement or by theoretical modelling, one has a complete description of the system's thermoelectric response, for any $V$ and $\Delta T$ small enough to remain in the linear response regime.
One can calculate the power output $P=-VI$ and the efficiency $\eta= P/J$ for any given (small enough) $\Delta T$ as a function of $V$.  
If one tunes $V$ to maximize the heat-engine's power output $P$ instead of $\eta$, one finds that the maximum power is 
\begin{eqnarray}
P_{\rm max}= \textstyle{{1 \over 4}} G S^2 \, \Delta T^2.
\label{Eq:P_max-linear}
\end{eqnarray}
With a bit more algebra, one can find the voltage $V$ at which the efficiency $\eta$ is maximal,
and prove that this maximal value of efficiency for a heat engine is
\begin{eqnarray}
\eta_{\rm max} = \eta_{\rm Carnot} \, {\sqrt{ZT+1} -1 \over  \sqrt{ZT+1} +1} \, ,
\label{Eq:eta_max_ZT}
\end{eqnarray}
where the Carnot efficiency in the linear response regime is given by $\eta_{\rm Carnot}=\Delta T /T$, and 
we define the dimensionless figure of merit 
\begin{eqnarray}
ZT = {GS^2T \over K}\, .
\label{Eq:ZT}
\end{eqnarray}
This dimensionless figure of merit is a simple measure of the quality of a thermoelectric; it is zero in the absence of  thermoelectricity, and we see from Eq.~(\ref{Eq:eta_max_ZT}) that Carnot efficiency requires that 
$ZT\to \infty$.
Current bulk semiconductor thermoelectrics have $ZT \sim 1$ (i.e.~maximum efficiency of about 
$\eta_{\rm Carnot}/6$), while it is commonly stated that they will become useful for everyday applications if one could get $ZT\sim3$ (i.e.~maximum efficiency of about 
$\eta_{\rm Carnot}/3$).  One sees immediately why the conduction of heat by phonons and photons is always bad in pure thermoelectric applications (see the next section); they contribute to $K$ the denominator of $ZT$ without making any contribution to the numerator.

As mentioned above, this linear response theory works well for bulk thermoelectrics, and the literature principally discusses maximizing $S$ and $ZT$. 
However, in nanostructures it fails as soon as $\Delta T/T$ (or $eV/k_{\rm B}T$) ceases to be small . When these are not small, 
$I$ and $J$ become highly non-linear functions of $V$ and $\Delta T$, which cannot be described in terms 
of $G$, $S$, etc.  As a result $ZT$ ceases to have any meaning, and instead the literature discusses power outputs and efficiencies.  
Thus, while modelling nanostructures in the linear response regime is extremely important for our understanding because it is often simpler than the non-linear regime, we believe that it is crucial to go beyond.  
We here mention an approach making use of symmetry relations of the time-evolution operator of open fermionic systems~\cite{Schulenborg2016Feb} that has recently been exploited to straightforwardly determine linear as well as nonlinear response-coefficients of weakly coupled quantum dots acting as thermoelectrics~\cite{Schulenborg2017Dec}.  
The rest of this review considers non-linear systems for which one must directly calculate the power outputs and efficiencies, and rarely mentions the linear-response quantities $S$ or $ZT$.

%%%%%%%%%%%%%%%%%%%%%%%%%%%%%%%%%%%%%%%%%%%%%%%%%%%%%%
\section{Parasitic heat flows: phonons and photons}
%%%%%%%%%%%%%%%%%%%%%%%%%%%%%%%%%%%%%%%%%%%%%%%%%%%%%%

A problem that is not to be overlooked in any quantum thermodynamic device, but which is particularly important
for nanostructures, is the problem of uncontrolled heat flows.  These are nearly always parasitic for the operation of the device.   Any device carries heat in the flow of electrons and also in the flow of photons and phonons.
The heat carried by the electrons can be controlled, e.g., by electric fields, and can be made to  produce electricity at relatively high efficiency.  However, the flow of photons and phonons is hard to control, 
and their flow from hot to cold does not generate any electric power.  
Thus, a device which efficiently converts the heat flow from hot to cold {\it carried by electrons} into electricity, 
will still have a very low thermodynamic efficiency if there is a larger heat flow from hot to cold
carried by {\it photons or phonons}.

The problem with phonons and photons is that they are hard to control.
As an example of this, it is intriguing to note (see Fig. 3 of Ref.~\cite{Benenti2017Jun}) that the materials which are the best thermal conductors (such as diamond or copper)
have a thermal conductivity which is only about 4 orders of magnitude higher than the worst heat conductors (glasses). Indeed, even the vacuum conducts heat through the exchange of photons in the form of 
black-body radiation.
One can compare this with electrical conductors, where the best (such as copper)
have an electrical conductance which is 24 orders of magnitude higher than the worst electrical conductors (such as diamond).  

Many nanostructures are studied in dilution refrigerators that cool them down to a fraction of a Kelvin
(the lowest achievable temperatures are of the order of 10 milli-Kelvins). At such low temperatures, the situation is less bad than at room temperature.  In particular, thermal photons in vacuum at less than a Kelvin have a wavelength of about a millimetre.  So two regions of smaller than millimetre size at different temperatures will have difficulty exchanging heat by thermal photons through vacuum, because they are smaller than a wavelength.   The situation is more complicated in nanostructured circuits, because the wires act as wave-guides that can carry a much shorter wavelength thermal photon from hot to cold \cite{Schmidt2004Jul,Pascal2011Mar}.
However, in general, phonons are a much more significant source of heat flow from hot to cold, since
thermal phonon wavelengths are typically tens of nanometres.
Most standard nanostructures are deposed on an electrically insulating substrate, which is a volume 
through which phonons can flow from the hot part of the nanostructure to the cold part.  
To avoid this, people are starting to  develop suspended nanostructures, where the substrate is replaced by a vacuum \cite{Bourgeois2005Feb,Zink2010Mar,Poran2017Mar,Cui2017Dec}.  This is a huge extra technical difficulty, but it may be essential to reduce phonon flows to acceptable levels.

Currently, the field of quantum thermodynamics has two big domains of interest. One domain is the subject of this review, and is to understand the transformation of heat into work in quantum systems (with the associated question of what is heat, work and entropy in such situations).
The second domain is to understand if and how isolated many-body quantum systems relax to a thermal state through internal interactions (with the associated question of how such a relaxation can be irreversible).
The above discussion of the difficulties of isolating a nanostructure from its environment should make it clear that it is hard to address the second domain experimentally with such nanostructures.   
Trapped atomic gases are better because it is easier to isolate the many-body system in question, than in nanostructures. 

%%%%%%%%%%%%%%%%%%%%%%%%%%%%%%%%%%%%%%%%%%%%%%%%%%%%%%
\subsection[]{Peltier cooling --- using the weakness of electron-phonon coupling at sub-Kelvin temperatures}
%%%%%%%%%%%%%%%%%%%%%%%%%%%%%%%%%%%%%%%%%%%%%%%%%%%%%%

An unusual feature of sub-Kelvin systems is that the electrons and phonons are much more weakly coupled to each other than at room temperature. The strength of the electron-phonon coupling goes like the 5th power of temperature, so the coupling is $10^{-12}$ times smaller at 1K than at 300K.  
Hence at low low enough temperatures,  one should think of the system as one containing two gases (one being the electrons and the other being the phonons) which both thermalize within themselves relatively fast, 
but which are so weakly interacting that they may be at different temperatures.
The dilution refrigerator cools the phonons (i.e. it cools the lattice), and the temperatures given for the temperature in the refrigerator are typically those of the phonons.  Experimentalists know that the electrons are often hotter, because the nanoscale circuit inside the refrigerator is coupled through wires and amplifiers to the electronics at room-temperature, used to control and probe the nanostructure.  
To ensure that the electrons in the nanostructure are as cold as the dilution refrigerator, one must make their coupling to room-temperature electrons much smaller than their coupling to the cold phonons in the dilution refrigerator.
Despite experimental progress in isolating the nanostructure from the room-temperature electronics,  this becomes increasingly challenging as one goes to lower temperatures, because of the weakness of the electron-phonon coupling.  

One idea of great interest is to use thermoelectric effects to cool the electrons directly,
thereby avoiding the electron-phonon coupling.  One could imagine a small gas of electrons in the dilution refrigerator (at some temperature close to the refrigerator temperature) being cooled to a much lower temperature through a thermoelectric effect.  Then the electrons would be colder than the phonons,
but the weakness of the electron-phonon coupling would mean that electrons would stay cold, because the phonons would be rather inefficient at heating them up.
Furthermore, one can also define non-invasive refrigerators, in the sense that they do not need to inject an electric current into the system to be cooled down~\cite{Partanen2016Feb,Sanchez2017Nov}, which can hence be further isolated from the rest of the circuit. 
In the long term it is hoped that such thermoelectric cooling could enable us to study the properties of electrons 
(new phases of matter, quantum phase transitions, etc.) at unprecedented  low temperatures. 
Most current experimental sub-Kelvin thermoelectric refrigerators rely on superconductors
which are discussed in Refs.~\cite{Giazotto2006Mar,Muhonen2012Mar,Courtois2016Dec}, although the refrigeration of microscopic semiconductor electron gases with quantum dots was also achieved some time ago~\cite{Prance2009Apr}.

%%%%%%%%%%%%%%%%%%%%%%%%%%%%%%%%%%%%%%%%%%%%%%%%%%%%%%
\subsection[]{Intrinsic leakage in electronic devices}
%%%%%%%%%%%%%%%%%%%%%%%%%%%%%%%%%%%%%%%%%%%%%%%%%%%%%%

In addition to the major problems of leakage heat currents carried by phonons and photons in otherwise electronic devices --- as described in the  sections above --- the electronic heat current itself can give rise to leakage. This is also known from bulk systems, where electrons can transfer heat by Coulomb interaction and electron-electron scattering, leading to electronic Fourier heat transfer in the absence of charge currents, see e.g. standard books like~\cite{Ashcroft1976Jan}. Its impact in nanoelectronic devices can easily be understood through the example of the quantum dot, shown in \textsf{Box 2}. When an electron tunnels into (or out of) the quantum dot in a single tunnelling process (sequential tunnelling), it takes along the energy determined by the  single-electron energy level of the quantized spectrum. However, when the quantum dot is strongly coupled to the reservoirs, higher-order tunnelling processes through energetically forbidden states (such as elastic and inelastic cotunnelling) can occur. These processes thereby allow for energy transfer through the dot in the absence of charge transfer. Furthermore, the possibly strong on-site Coulomb interaction between electrons on the dot means that the energy to be paid in the transition $1\leftrightarrow2$ is bigger by $U$ with respect to the energy to be paid at the transition $0\leftrightarrow1$. A sequence of tunnelling processes through the dot therefore allows for the transfer of the interaction energy $U$, while no charge current is flowing.

%===================================================
\begin{table}[b]
\fbox{\hskip 3mm
\begin{minipage}{0.9\textwidth}
\sffamily
\vskip 2mm
\textbf{BOX 2: QUANTUM DOTS}
\vskip 2mm
\justify
Quantum dots can be experimentally realized in a large variety of ways, including the patterning of semiconductor heterostructures, etching graphene, contacting carbon nanotubes or other types of molecules, etc. See e.g. Refs.~\cite{Kouwenhoven2001May,Hanson2007Oct} or text books in which these devices are treated~\cite{Bird2003,Ihn2009Dec}. In a quantum dot, electrons are confined and occupy discrete energy levels, similar to the situation in an atom. Furthermore, due to the smallness of the device, electrons are subject to strong Coulomb interaction. One of the important features of a quantum dot is that it can be contacted to electronic reservoirs (allowing for electronic transport through it) and controlled by external gates to which voltages can be applied. While there is possibly a large number of electrons in the quantum dot, the physics of the device is typically governed by tunnelling processes of single electrons. When describing the dot physics, we therefore set the reference occupation number to '0' and from here on talk about occupation states '0,1,2' referring to \textit{extra-electrons} that enter the dot due to the application of bias or gate voltages. \\

\noindent
\begin{minipage}[c]{6.3cm}
\justify
$\ $ The figure shows a sketch of the energy landscape of a quantum dot in contact with two electronic reservoirs. In this simplified (however experimentally relevant example!) we assume that a single electronic level is energetically accessible; it has the energy $\varepsilon$, which is tunable via an external gate through an additional term proportional to $-eV_\text{g}$. Such a discrete energy level is an optimal realization of the peaked energy spectrum useful to improve thermoelectric operation~\cite{Mahan1996Jul}. 
\end{minipage}
\hspace{0.1cm}
\begin{minipage}[r]{8.2cm}
\includegraphics[width=7.4cm]{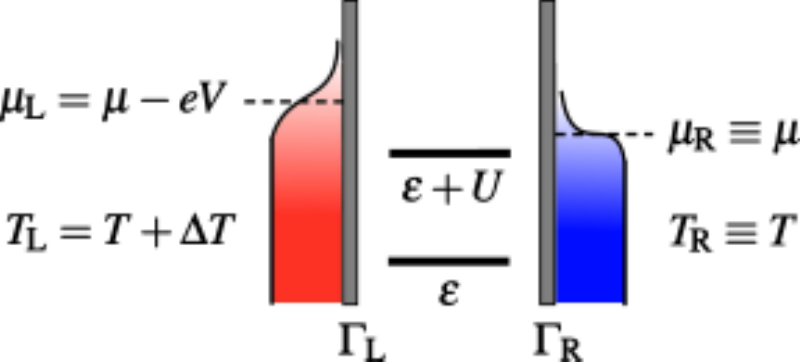}\vspace{0.4cm}\\
\end{minipage}\\

In order to add a second electron on the dot, the Coulomb interaction energy $U$ has to be paid. The coupling to the left and right reservoirs is characterized by the coupling strengths $\Gamma_\text{L}$ and $\Gamma_\text{R}$. In contrast to the quantum dot, the electronic reservoirs have a continuous spectrum and their occupations are given by equilibrium Fermi-functions with respect to a well-defined temperature and electrochemical potential, which can be different in each reservoir. In this example, we show Fermi distributions of a cold and a hot reservoir, relevant in the context of thermoelectrics. Note that the temperature \textit{in the dot} is typically not a well-defined quantity!

$\ $

\end{minipage}\hskip 3mm}
\end{table}
%======================

%%%%%%%%%%%%%%%%%%%%%%%%%%%%%%%%%%%%%%%%%%%%%%%%%%%%%%%%%%%%%%%%%%%%%%%%%%%%%%%%%%%%%%%%%%%%%%%%%
\section{Multi-terminal steady-state machines with quantum dots}
\label{sec:multi_terminal}
%%%%%%%%%%%%%%%%%%%%%%%%%%%%%%%%%%%%%%%%%%%%%%%%%%%%%%%%%%%%%%%%%%%%%%%%%%%%%%%%%%%%%%%%%%%%%%%%%

In this section, we discuss examples for implementations of heat engines using quantum dots embedded in a multi-terminal electronic set-up. A quantum dot, see  \textsf{Box 2}, is characterized by a discrete energy-level spectrum (similar to an atom) and possibly strong on-site Coulomb interaction. Quantum dots can be tunnel-coupled to electronic reservoirs allowing for electronic transport through them and their properties, such as the level spectrum, can be tuned via the application of external gates.
These and other properties make electronic devices with quantum dots interesting for thermoelectrics or for nanoscale heat engines:  first of all, their discrete spectrum provides a means for \textit{energy-selective transport}, which is of high relevance for the efficiency of thermoelectric applications~\cite{Hicks1993May,Hicks1993Jun,Mahan1996Jul,Humphrey2005Mar}. Strong Coulomb interaction between electrons on different, \textit{purely capacitively} coupled quantum dots can serve for the transfer of a well-defined amount of energy from a heat bath to a thermoelectrically active region. Finally, the idea of using a quantum dot --- characterized by a small number of electronic states --- as working substance of a thermodynamic heat engine, requires a completely new understanding of these devices. The reason for this is that, historically, heat engines are made of large systems to which a statistical description applies. This however obviously breaks down in few-level devices such as quantum dots.
Indeed, we will see at the end of this section that the resulting \textit{absence of thermalization} in quantum dots in a non-equilibrium set-up can lead to counter-intuitive effects. 
All this makes quantum dot heat engines interesting from a fundamental point of view; however, as mentioned before, this research is also guided by a strong need for finding new ways of on-chip refrigeration and waste-heat recovery for future nanoelectronic applications, where quantum dot devices constitute fundamental building blocks. 

%%%%%%%%%%%%%%%%%
\begin{figure}[t]
\begin{minipage}{0.44\textwidth}
\justify
(a)
\vskip -4mm
\hskip 4mm 
\includegraphics[width = 0.95\textwidth]{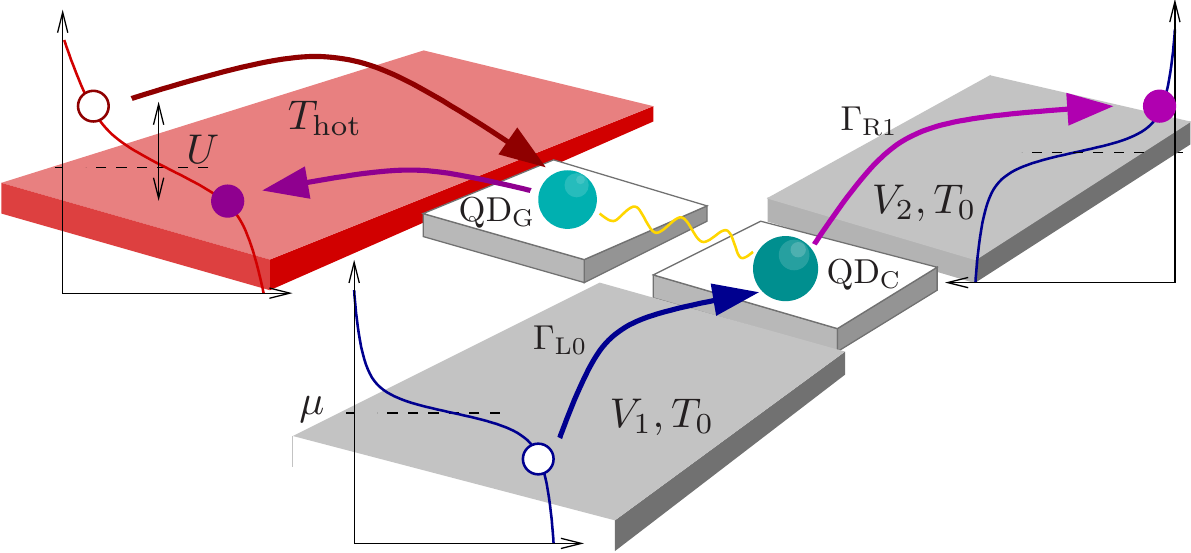}\\
\vskip 8mm
(b)
\vskip -3mm
\hskip 5mm 
\includegraphics[width = 0.8\textwidth]{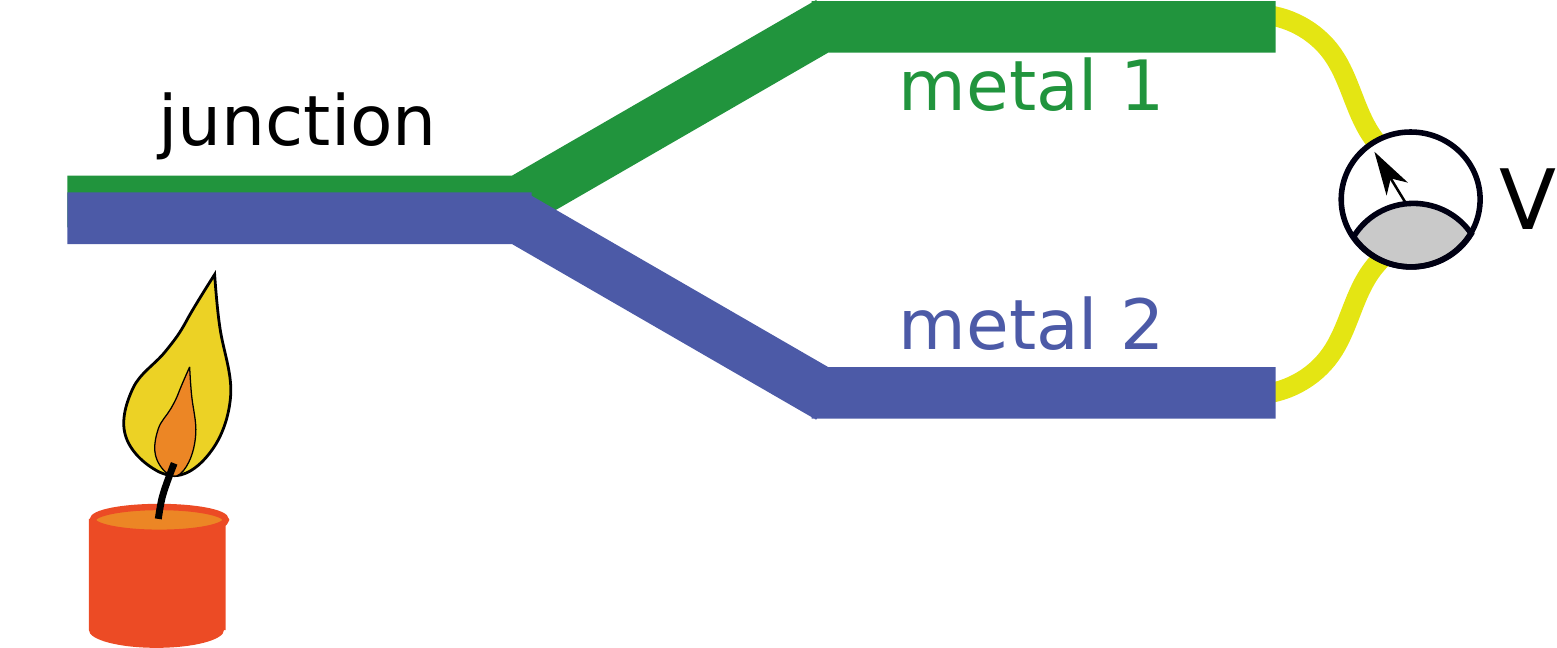}
\end{minipage} 
\hskip 2mm
\begin{minipage}{0.44\textwidth}
\justify
(c)
\vskip -4mm
\hskip 3mm 
\includegraphics[width=\textwidth]{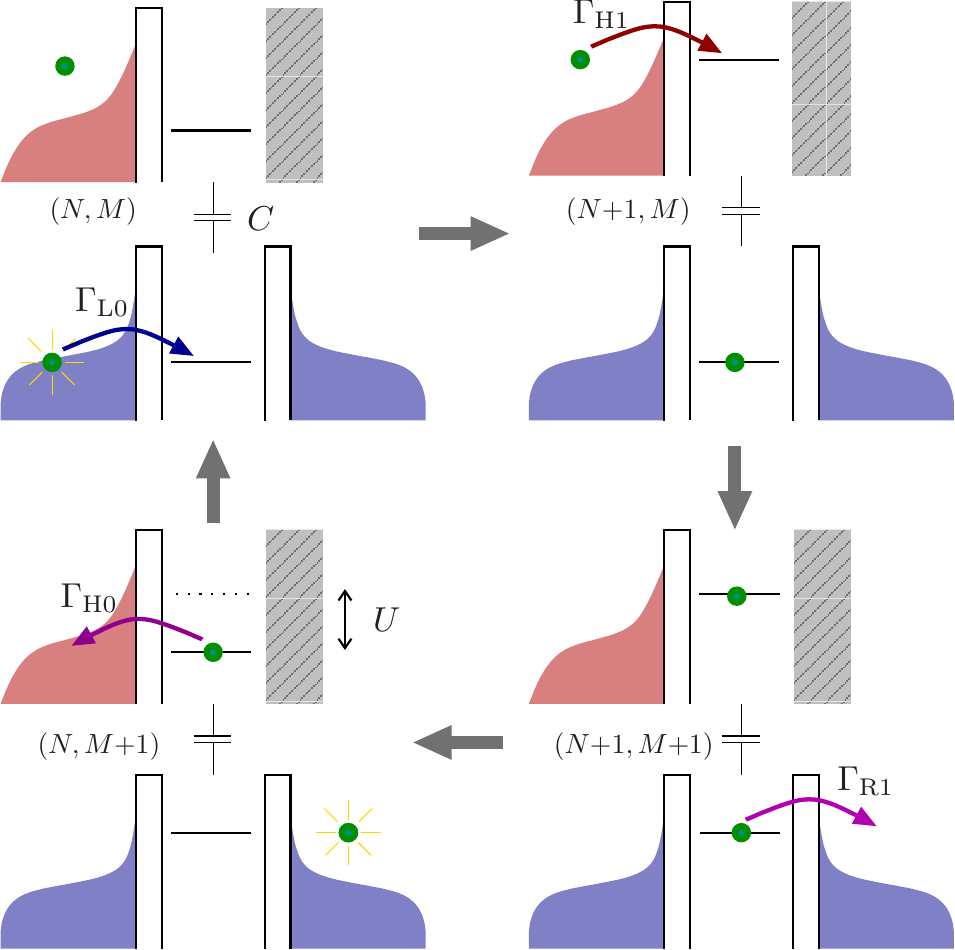}
\end{minipage}
\caption{\label{fig:three_terminal} 
(a) Sketch of the three-terminal double-dot set-up acting as a nanoscale thermocouple. Heat is transferred from the electronic heat bath (in red)  to the working substance (consisting in the quantum dot QD$_\text{C}$ tunnel coupled to source and drain reservoir) via capacitive coupling between the two dots QD$_\text{G}$ and QD$_\text{C}$. Coloured arrows, together with indicated states on the reservoirs Fermi distribution indicate the process shown in detail in (c).
(b) Sketch of a classical thermocouple. (c) The working principle of the nanoscale thermocouple, shown as a thermodynamic cycle.  The red and blue distributions are the Fermi functions of the hot and cold reservoirs respectively,
The vertical axis is energy, while the horizontal axis in the distributions is the probability, with which each energy state is occupied. (i.e.\ it shows a more quantitative representation of the electron distributions than sketched in Fig.~\ref{fig:energy-filter}). Grey, hatched regions indicate a boundary of the quantum dot without tunnel coupling.
}
\end{figure}
%%%%%%%%%%%%%%%%%%

Here we present a quantum-dot equivalent of a thermocouple (or energy harvester), proposed and analysed in Ref.~\cite{Sanchez2011Feb},  and recently realized experimentally in different groups~\cite{Thierschmann2015Aug,Roche2015Apr}, see Fig.~\ref{fig:three_terminal}(a) for a sketch of the set-up.  One of the advantages of the multi-terminal design is that the heat bath is spatially separated from the working device.  This might for example result in advantages with respect to the leakage effects due to phonons or tunnelling as discussed before. The set-up basically consists of two parts: the active thermoelectric region, shown in the lower, right part of the sketch is made of the central quantum dot, QD$_\text{C}$, tunnel-coupled to two electronic contacts at the same temperature $T_0<T_\text{hot}$. The upper, left part of the sketch shows the heat bath in contact with a second dot QD$_\text{G}$. The occupation of this second dot  strongly fluctuates due to the large temperature, $T_\text{hot}$, of the hot electrode, constituting the heat bath. This dot QD$_\text{G}$ is \textit{purely capacitively} coupled to the working substance, depicted in the lower half of the sketch. It is via this capacitive coupling between quantum dots (namely due to Coulomb interaction) that heat transfer between the heat bath and the thermoelectrically active region takes place. For simplicity, we now assume that each of the two dots can at most accept one extra-electron due to strong on-site Coulomb interaction. The two dots - forming a purely capacitively coupled double dot - can however each be singly occupied at the same time (such that the double dot is doubly occupied); in this case the Coulomb interaction $U$ needs to be provided.
The working principle of this quantum-dot thermocouple can be understood by following the processes depicted in the sketch in Fig.~\ref{fig:three_terminal}(c) and also alluded to by the arrows in Fig.~\ref{fig:three_terminal}(a). Assume that the two dots are initially empty and that - due to fluctuations - the lower dot gets occupied by an electron from the left reservoir. This electron has to be at the required single-particle energy for the transition from an empty to a single-occupied dot. If now a further electron enters the double dot from the hot reservoir by tunnelling into QD$_\text{G}$, an additional energy $U$ has to be paid due to the capacitive coupling to the \textit{occupied} lower dot. This extra energy $U$ can be transferred to the working substance, when in the next process the electron of the lower dot QD$_\text{C}$ tunnels out again, for example to the right reservoir. The described process is  one of many possible tunnelling sequences that can in general occur in such a set-up. However, if $T_\text{hot}>T_0$, such that heat is on average transferred from the upper to the lower part of the device and if at the same time the asymmetry condition $\Lambda\neq0$ is fulfilled for
\begin{equation}
\label{eq:asymmetry}
\Lambda=\frac{\Gamma_{\text{L}0}\Gamma_{\text{R}1}-\Gamma_{\text{L}1}\Gamma_{\text{R}0}}{\left(\Gamma_{\text{L}0}+\Gamma_{\text{R}0}\right)\left(\Gamma_{\text{L}1}+\Gamma_{\text{R}1}\right)}\ ,
\end{equation}
the described type of process can actually be used in order to do electrical work. The feasibility of this scheme has been proven in experiments~\cite{Roche2015Apr,Hartmann2015Apr,Thierschmann2015Aug}. Here, work is done by creating a \textit{directed current against a bias voltage} in the lower part of the set-up - the working substance.  The condition $\Lambda\neq0$ requires an asymmetry both in real space (the tunnel-coupling strength to the left reservoir, $\Gamma_\text{L}$ has to be different from the tunnel coupling strength  to the right reservoir, $\Gamma_\text{R}$) but also an asymmetry in the energy-dependence of these couplings has to occur. The subscripts 0 and 1 indicate the tunnel couplings for the case that the upper dot is empty (0) or filled (1). Physically, such an asymmetry would make the process of Fig.~\ref{fig:three_terminal}(c) more probable than the ones with reversed directions or reversed order of tunnelling events. For the specific example, it would mean that it is more probable to fill the empty dot QD$_\text{C}$ from the left reservoir when the upper dot QD$_\text{G}$ is empty ($\Gamma_{\text{L}0}>\Gamma_{\text{R}0}$), while it is more probable to empty the dot QD$_\text{C}$ by an electron tunnelling to the right reservoir, when the upper dot QD$_\text{G}$ is filled ($\Gamma_{\text{R}1}>\Gamma_{\text{L}1}$).

The resulting charge current, $I$,  between the left and the right contact is proportional to the heat current, $J$, transferred from the heat bath into the working substance \textit{via the Coulomb interaction energy}. In the absence of a bias voltage between left and right reservoir, the charge current takes the simple form 
\begin{equation}
\label{eq:thermocouple}
I=\frac{-e\Lambda}{U}J_{\rm G}.
\end{equation}
If the asymmetry factor takes its maximal value, namely if it is equal to 1, then exactly one electron is transferred from left to right with each transfer of energy $U$ from the heat bath to the working substance. Note that the asymmetry given in Eq.~(\ref{eq:asymmetry}) constitutes the analogue of the two \textit{dissimilar} metals of a classical thermocouple, as sketched in Fig.~\ref{fig:three_terminal}(b).

%%%%%%%%%%%%%%%%%
\begin{figure}[t]
\justify
(a) \hskip 0.45\textwidth (b)
\vskip -4mm\hskip 5mm
\includegraphics[width = 0.42\textwidth]{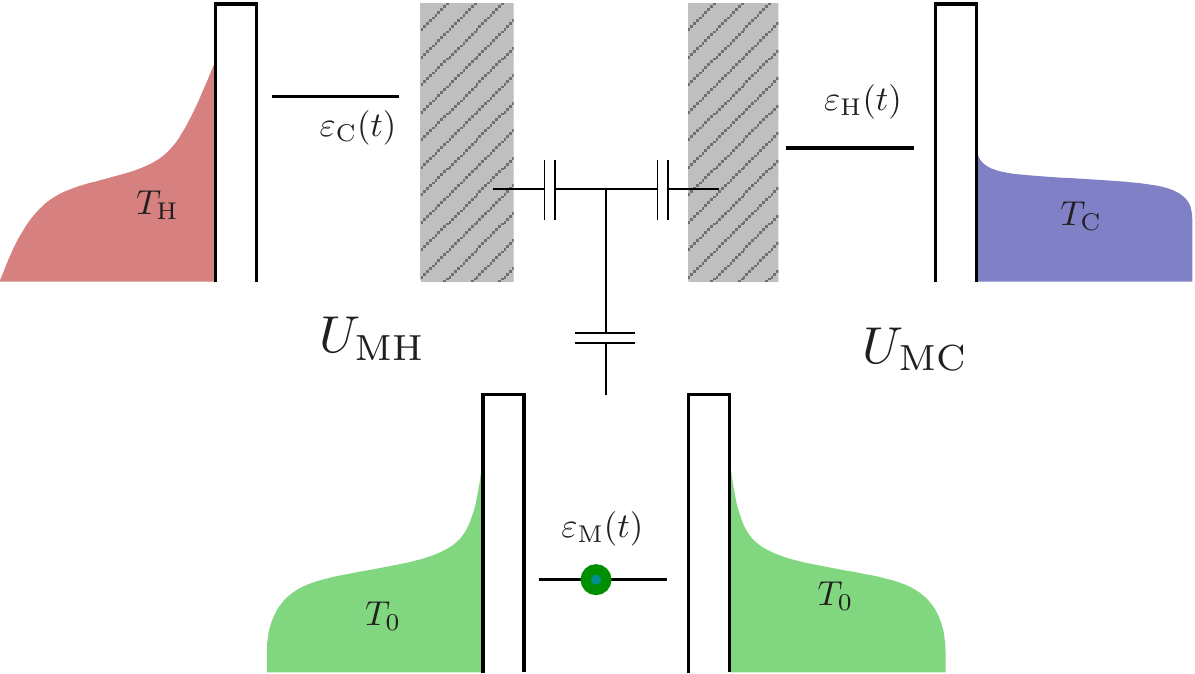}

\vskip -4.cm
\hskip 0.5\textwidth
\includegraphics[width = 0.45\textwidth]{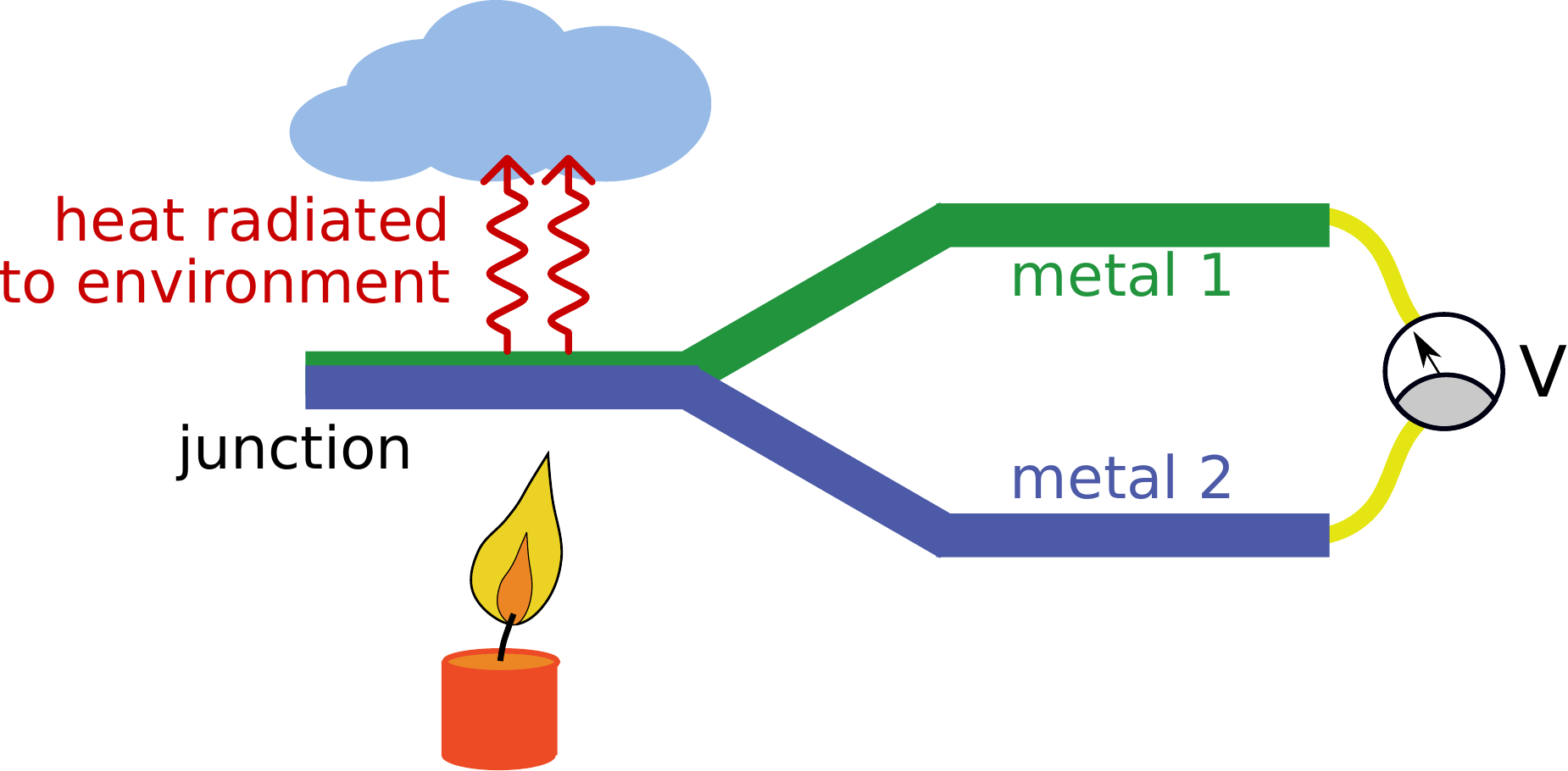}
\vskip 1cm
\caption{\label{fig:four_terminal} 
(a) Sketch of the four-terminal triple-dot set-up acting as a nanoscale thermocouple with two heat baths. Compared to the set-up shown in Fig.~\ref{fig:three_terminal}, the working substance is now additionally coupled to a cold bath. Also in this case the heat transfer between cold bath and working substance takes place via capacitive coupling between two dots. Note that in general, also capacitive coupling between the two upper dots can occur, which is detrimental for the operation of this device, but does not hinder the observed effects to occur. (b) Classical analogue, where the cold heat bath could be given by the environment to which the thermocouple can radiate heat. 
}
\end{figure}
%%%%%%%%%%%%%%%%%%

As discussed previously, a big difference between classical heat engines and their quantum dot analogues lies in their dimensions with respect to the thermalization length. While a well-defined temperature can be associated to the bimetallic part of the classical thermocouple, this is in general not the case for a quantum dot when it takes the role of the working substance of a thermocouple as described above. This becomes particularly clear when one pushes the analogy to the classical thermocouple even further and assumes that in addition to the heat bath (indicated by the candle in the sketch in Figs.~\ref{fig:three_terminal}(c) and~\ref{fig:four_terminal}(b)) the thermocouple is in thermal contact also with a colder heat bath, e.g. the environment (indicated by the cloud in ~\ref{fig:four_terminal}(b)). Also this set-up can be mimicked by quantum dots as shown in Fig.~\ref{fig:four_terminal}(a), see Ref.~\cite{Whitney2016Jan}.

Since now there is a capacitive coupling, $U_\text{MH}$ and $U_\text{MC}$, between the working substance (dot with reservoirs) and \textit{both} the additional quantum dots in contact with a hot  and a cold reservoir,  the total induced charge current (in the absence of voltage differences),\footnote{This simple relation is valid only if the direct capacitive coupling between the upper dots (which tunnel-couple to the hot and cold reservoirs) is so strong that they cannot be occupied simultaneously. Note however, that the  effects described here continue to exist also in the presence of a smaller capacitive coupling, $U_\text{HC}\sim U_\text{MH},U_\text{MC}$.}
\begin{equation}
\label{eq:thermocouple2}
I=-\frac{e\Lambda_\text{C}}{U_\text{MC}}J_\text{C}-\frac{e\Lambda_\text{H}}{U_\text{MH}}J_\text{H}.
\end{equation}
has a contribution proportional to the heat current from the hot reservoir $J_\text{H}$ to the working substance and the heat current from the cold reservoir, $J_\text{C}$, to the working substance. Here,  $\Lambda_\text{C}$ and $\Lambda_\text{H}$ are the asymmetry factors in analogy with Eq.~\eqref{eq:asymmetry} with respect to the occupation of the dot in contact with the cold or the hot reservoir. This simple extension of the previous results, Eqs.~\eqref{eq:asymmetry} and \eqref{eq:thermocouple}, leads to an interesting twist: Even when the total heat current flowing from the heat baths \textit{into} the working device, $J_\text{in}=J_\text{H}+J_\text{C}$, vanishes, a charge current can still be induced by the heat current flowing \textit{through} the device from the hot into the cold reservoir, $J_\text{trans}=J_\text{H}-J_\text{C}$. This statement even holds when the induced current is transported against a voltage gradient between the left and right reservoir, thereby doing work. This observation leads to a number of seemingly paradoxical observations.
The first direct consequence of this is obviously the possibility of doing work without absorbing heat from a heat bath! This means in particular that, in order to maintain energy conservation when doing work, heat has to be extracted from the reservoirs of the working substance, thereby cooling it down! Note that these observations - even though highly counter-intuitive - do however, as required, not violate any of the laws of thermodynamics. Not only is the total energy conserved in the process. Also the entropy of the total system does indeed increase while work is done. This is because the heat flow from the hot to the cold reservoir, which \textit{drags} the induced current without any energy transfer, leads to an entropy production.

These effects result from the inability of the quantum dot to reach a thermal equilibrium.
Importantly, when forcing the quantum dot to thermalize (for example via an additional probe contact attached to the central dot), the counter-intuitive power production without heat absorption is suppressed and the behaviour expected from a classical thermocouple is restored.

%===================================================
\begin{table}[b]
\fbox{\hskip 3mm
\begin{minipage}{0.9\textwidth}
\sffamily
\vskip 2mm
\textbf{BOX 3: QUANTUM HALL EFFECT}
\vskip 2mm
\justify
A paradigmatic case of quantum electronic transport is the quantum Hall effect~\cite{Klitzing1980Aug}. The occurrence of quantized conductance plateaux in this regime is a striking evidence for the fact that precise quantum measurements are not restricted to atomic scales but can also be performed in massive, disordered materials. This is ultimately due the quantization of cyclotron orbits (Landau levels) that lie below the\\ 
\begin{minipage}{0.55\textwidth}
\justify
 Fermi energy in the centre of the sample but overcome it close to its borders, thereby forming chiral edge states~\cite{Halperin1982Feb}. 
Electron propagation in edge states occurs in analogy to photon propagation in a waveguide.  However, electronic propagation in edge states is protected from backscattering by the chirality of the sample~\cite{Buttiker1988Nov}, even in the presence of (moderate)  impurities and disorder (unlike for photons in waveguides). 
The figure on the right shows a typical Hall set-up, where a current is injected between terminals 4 and 2 and the transverse voltage is measured between probe terminals 1 and 3. Since electrons flow chirally along the edges of the sample, with its direction determined by the applied magnetic field, all electrons injected from terminal 4 are hence absorbed by terminal 1. 
\end{minipage}\hspace{0.2cm}
\begin{minipage}{0.45\textwidth}
\vskip 3mm
\includegraphics[width = 0.89\textwidth]{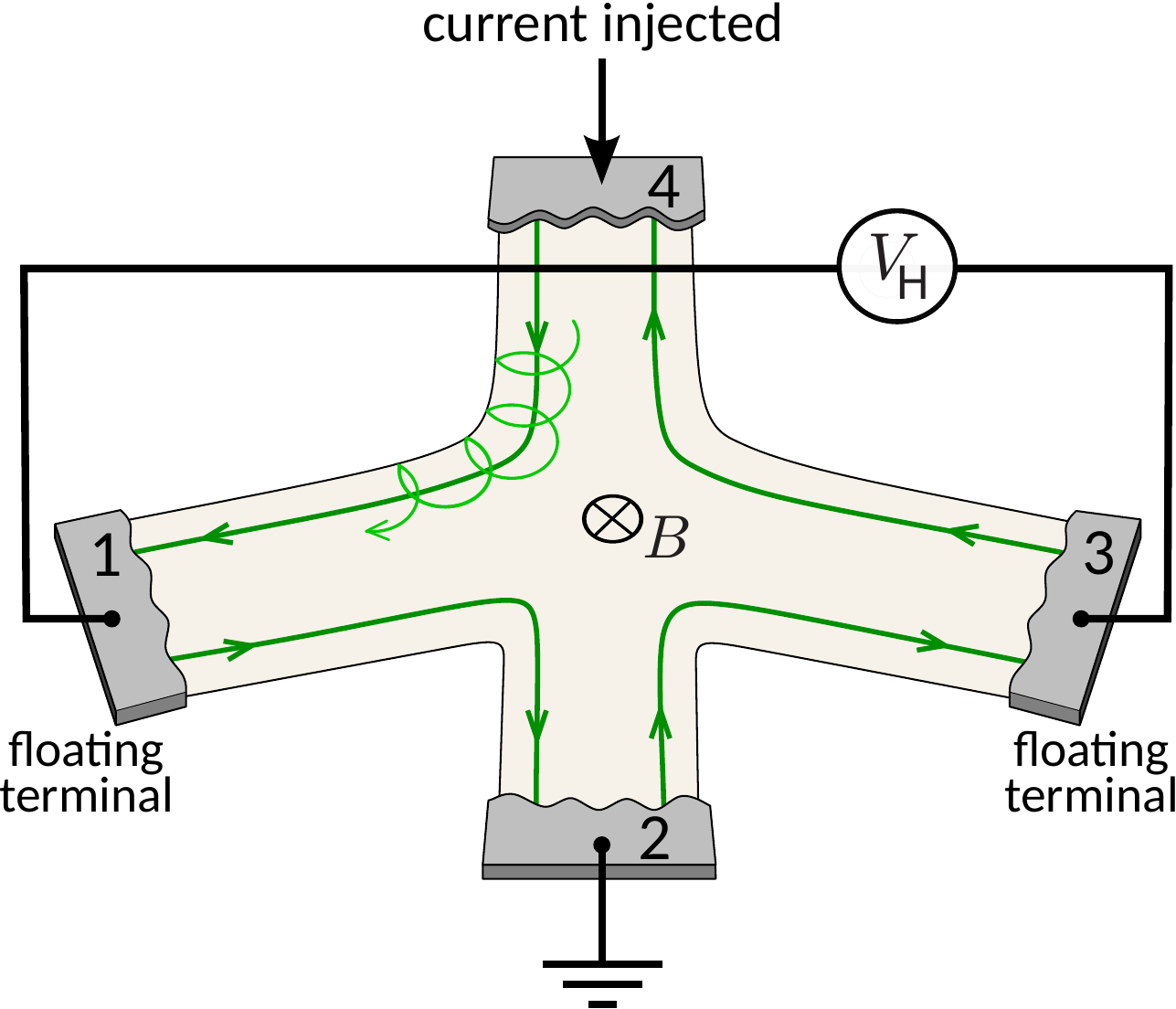}
\end{minipage}\\
\vskip 3mm
\begin{minipage}{0.25\textwidth}
\includegraphics[width=\textwidth]{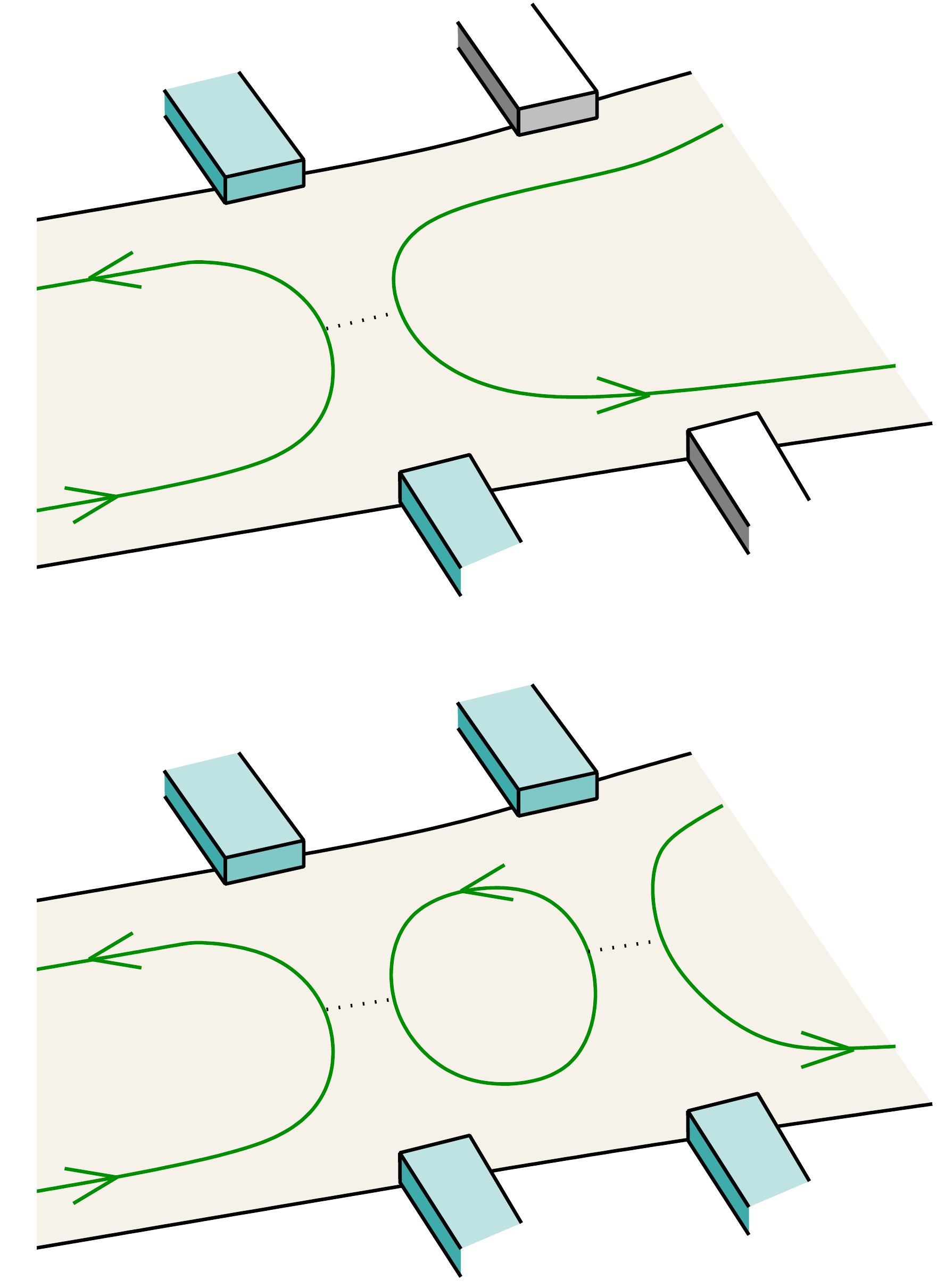}
\end{minipage}\hspace{0.2cm}
\begin{minipage}{0.7\textwidth}
\justify
 Since, furthermore, probes do not inject a net current, its potential must adapt to re-emit the same amount of charge into terminal 2. The same can be said of terminal 3, leading to a finite Hall voltage $V_{\rm H}=V_1-V_3=(h/e^2)I$.  The associated resistance (the von Klitzing resistance) is quantized and determined by fundamental constants (the charge of the electron, $e$, and Planck's constant, $h$), and eventually by the number of channels, $N$: $R_{\rm K}=h/(Ne^2)$.

The application of lateral gate voltages to the Hall bar can bring the two edges close enough to each other, such that electrons can be backscattered by being transferred from one channel to the opposite. This means that a point contact  is formed (in analogy to beam splitters in optics), see left figure for one (respectively two) pairs of gates to which a voltage is applied.
This permits to perform scattering experiments on quantum point contacts~\cite{Wees1988Feb}, or even on different types of interferometers, like the Fabry-Perot interferometer shown in the bottom panel. 

$\ $
\end{minipage}
\end{minipage}\hskip 3mm}
\end{table}
%======================

%%%%%%%%%%%%%%%%%%%%%%%%%%%%%%%%%%%%%%%%%%%%%%%%%%%%%%%%%%%%%%%%%%%%%%%%%%%%%%%%%%%%%%%%%%%%%%%%%
\section{Quantum Hall thermoelectrics}
%%%%%%%%%%%%%%%%%%%%%%%%%%%%%%%%%%%%%%%%%%%%%%%%%%%%%%%%%%%%%%%%%%%%%%%%%%%%%%%%%%%%%%%%%%%%%%%%%

Quantum Hall systems, in which electron transport takes place along chiral edges, have several advantages from the thermoelectric point of view. 
Firstly, in quantum Hall devices, simple quantum point contacts are particularly easy to make, and allow for the implementation of thermoelectric effects. A quantum point contact forms a narrow constriction whose width controls the number of channels that can be transmitted, see also  \textsf{Box 3}. Different channels can be totally transmitted, totally reflected, or partially transmitting. The partially transmitting  ones are of interest for thermoelectrics, because the transmission probability  is energy-dependent~\cite{Buttiker1990Apr} (rapidly switching from low to high transmission with the magnitude of the electron's energy). This upper-pass filter is enough to break electron-hole symmetry and to give a finite thermoelectric response~\cite{Whitney2014Apr,Whitney2015Mar}. Quantum point contacts were used for the first thermoelectric experiments in mesoscopic systems~\cite{Streda1989,Molenkamp1990Aug,Molenkamp1992Jun}. It has recently even permitted to infer the quantum of thermal conductance of a single \textit{electronic} conduction channel~\cite{Jezouin2013Nov}, related to Eq.~\eqref{maxJ}.

Secondly, transport through the sample is phase coherent up to device dimensions of the order of $\mu$m. This yields the possibility to have the heat source at a relatively large distance from the system of interest acting as the working substance. Thereby, the latter can be kept cold enough to perform operations. In this way, larger temperature gradients can be applied without overheating the system.  
Due to the long phase coherence length, it is furthermore possible to construct different types of interferometers (in analogy to optics) using quantum point contacts as beam splitters. The thermoelectric response in these interferometers is then possible uniquely due to quantum interference occurring thanks to the energy dependence of the phase of the electronic wave function, which is gained in the propagation between beamsplitters~\cite{Hofer2015May,Vannucci2015Aug,Samuelsson2017Jun}.

%%%%%%%%%%%%%%%%%%%%%%%%%%%%%%%%
\begin{figure}[t]
%\begin{center}
\includegraphics[width = 0.9\textwidth]{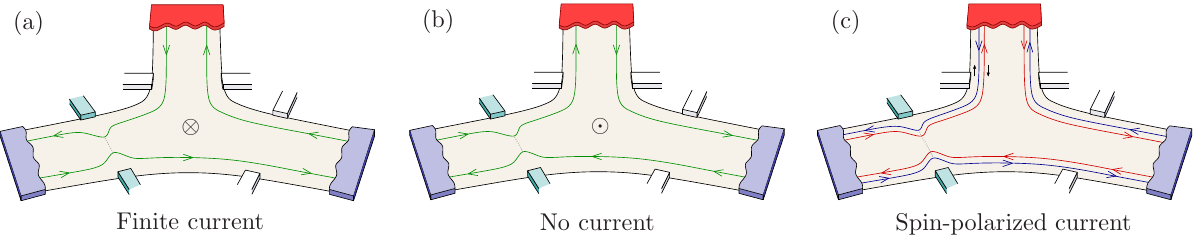}
\caption{\label{fig:3termqhe} 
Chiral thermoelectrics in the quantum Hall regime. We show a three-terminal set-up, in which the upper terminal acts as a heat bath injecting a heat current into the lower part of the set-up.  Only a single quantum point contact is required to produce a thermoelectric response by conversion of a heat current injected from terminal 3 into a charge current between terminals 1 and 2. In (a) electron-hole pairs injected from the hot reservoir impinge on the quantum point contact. In (b) the propagation direction of the edge channels is inverted by the inversion of the magnetic field. The result is that electron-hole pairs thermalize without having passed through the quantum point contact. (c) Modified set-up based on a topological insulator, in which spin-polarized edge states arise in the absence of a magnetic field. In all three panels, gate electrodes coloured in green are assumed to form a quantum point contact by application of a gate voltage, while we assume that no voltage is applied to gate electrodes coloured in white.
}
%\end{center}
\end{figure}
%%%%%%%%%%%%%%%%%%%%%%%%%%%%%%%%

In this section, we will go further into a third reason making quantum Hall devices interesting for thermoelectrics. This is the straightforward possibility to design non-trivial multi-terminal devices --- the quantum Hall effect being a multi-terminal effect by its own nature. 
This makes it possible to  introduce quantum analogues of the Nernst effect~\cite{Stark2014Apr,Sothmann2014Aug}, where the injection of a heat current leads to a \textit{transverse} electric response.\footnote{The thermoelectric response measured in a series of terminals has been used to probe how energy relaxation takes place along the edge, involving an elegant demonstration of chirality~\cite{Granger2009Feb,Nam2013May}. }
A minimal configuration with three terminals, as shown in Fig.~\ref{fig:3termqhe}, exhibits a peculiar behaviour as compared with a \textit{trivial} three-terminal thermocouple. Let us consider a central probe terminal (indicated in red in all three figure panels) which injects heat into a system which is otherwise at the same lower temperature~\cite{Jordan2013Feb,Donsa2014Mar,Whitney2016May} (indicated by the two reservoirs at the bottom of the set-up  depicted in blue). We now assume that there is a constriction realized by a quantum point contact on one side of the system only, see again Fig.~\ref{fig:3termqhe}.  Electron-hole excitations created at the hot terminal and impinging on the lower part of the device are separated at the quantum point contact, when the latter is placed in the direction of their propagation,  as realized in the sketch in panel~\ref{fig:3termqhe}(a). This separation of electron and hole excitations results in a thermoelectric effect, namely a charge current is induced between the two lower contacts (respectively a thermovoltage builds up if the contacts are floating).

However, as soon as the magnetic field is inverted, the propagation direction along the edges is inverted, see panel (b) of Fig.~\ref{fig:3termqhe}. In that case, non-equilibrium electron-hole excitations propagate in the opposite direction and are all absorbed and thermalized at the opposite terminal~\cite{Sanchez2015Apr,Sanchez2015Jul}. As hot electrons and holes are never separated at the junction, no thermovoltage is generated. The chirality of the quantum Hall edge states thus manifests in the absence of a thermoelectric effect if one reverses the magnetic field. In this way, thermoelectrics can been used to probe the presence of chiral states. 
Interestingly, this absence of a Seebeck response (a thermovoltage) in the configuration of Fig.~\ref{fig:3termqhe}(b) comes with a finite Peltier response (namely, the application of a voltage can be used for cooling). This is a remarkable situation. As discussed above, the Seebeck and Peltier coefficients are directly proportional to each other whenever the system is time-reversal invariant~\cite{Onsager1931Feb}. A magnetic field obviously breaks this symmetry~\cite{Buttiker1988May,Butcher1990}, however in general both the Seebeck and the Peltier coefficient vanish simultaneously. To the best of our knowledge, only in the fully chiral system, realized in the quantum Hall regime, one of them can be zero while the other one is finite. What is remarkable, is that by changing the sign of the magnetic field, one can simply chose which of the two coefficients is zero and which one leads to a nonvanishing thermoelectric effect.

We finally want to mention two interesting extensions of this simple set-up. First, such an analogue of the Nernst effect can be used to probe the presence of edge states in two-dimensional topological insulators. In topological insulators, electrons with opposite spin flow in opposite directions due to a strong spin-orbit coupling. Transport is then said to be helical. 
In the geometry shown in Fig.~\ref{fig:3termqhe}(c), one can see that spin up and down electrons propagate in opposite directions along the same edge. This picture can be immediately seen as a combination of the situation in  Figs.~\ref{fig:3termqhe}(a) and (b) but for opposite spins. Namely, only one of the channels (the spin-up one, in this case) will lead to a thermoelectric effect, when a temperature gradient is applied between the upper contact and the lower contacts, as discussed above. Hence, the generated current is fully spin-polarized~\cite{Sanchez2015Apr}.  By optionally applying a gate voltage to the quantum point contacts on the two sides, the spin-polarization of the current induced by the temperature gradient can be controlled. Such effects are of interest in the context of spintronics~\cite{Mani2018Feb}. 

As a second extension of the  set-up discussed above, the broken time-reversal symmetry induced by the magnetic field has been suggested as a means to increase the thermoelectric efficiency at the point of maximum power generation~\cite{Benenti2011Jun}. In general, the upper bounds for the thermoelectric efficiency can be obtained from symmetry arguments~\cite{Brandner2013Feb}. A recently studied example is an engine based on the (four-terminal) Nernst effect, where the efficiency bounds can be reached only in the quantum regime~\cite{Sothmann2014Aug}, thereby outperforming the classical version~\cite{Stark2014Apr}. The use of 
fractionally-charged carriers in fractional quantum Hall states (states induced by the interplay of electron-electron interactions and strong magnetic fields) has also been suggested in order to improve the efficiency~\cite{Roura-Bas2018Feb}.

%%%%%%%%%%%%%%%%%%%%%%%%%%%%%%%%%%%%%%%%%%%%%%%%%%%%%%%%%%%%%%%%%%%%%%%%%%%%%%%%%%
\section{Beyond steady state: Heat engines with a time-dependent cycle}
%%%%%%%%%%%%%%%%%%%%%%%%%%%%%%%%%%%%%%%%%%%%%%%%%%%%%%%%%%%%%%%%%%%%%%%%%%%%%%%%%%

The tunability of the properties of small electronic structures, for example by applying voltages to gate electrodes, makes them interesting for the realization of heat engines with a cyclic operation. Various examples have been studied in recent years, see e.g. Refs.~\cite{Arrachea2007Jun,Ludovico2016Feb,Bruch2017Dec,Calvo2017Oct}, some of them relying on quantum interference~\cite{Karimi2016Nov}. Also in experiments, nanoelectronic cyclic heat engines have been realized~\cite{Koski2014Sep,Koski2015Dec}.

\begin{figure}[t]
%\begin{center}
\justify
(a) \hskip 0.42\textwidth (b) \hskip 0.23\textwidth (c)

\begin{minipage}{ 0.41\textwidth}
\includegraphics[width = \textwidth]{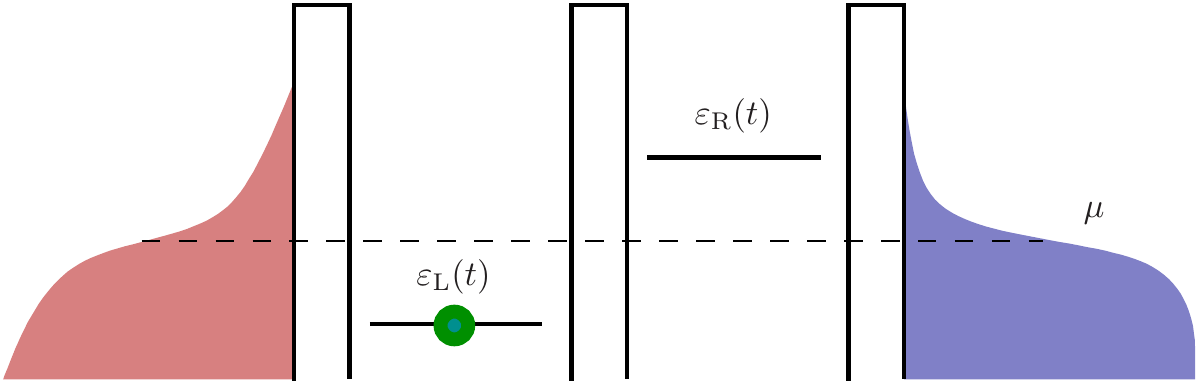}
\end{minipage}\hspace{0.7cm}
\begin{minipage}{0.5\textwidth}
\includegraphics[width = \textwidth]{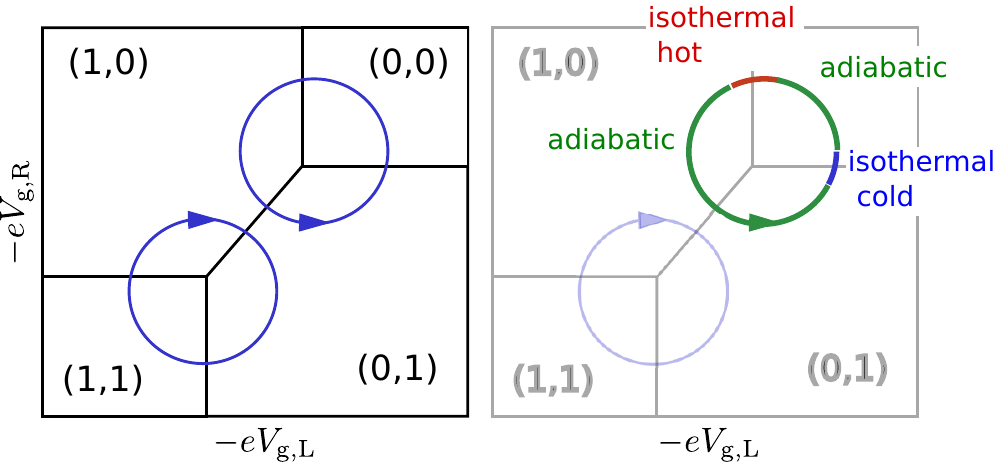}
\end{minipage}
\caption{\label{fig:cyclic} 
(a) Energy landscape of  a double quantum dot, weakly coupled to a left and right reservoir which can be at different temperatures and electrochemical potentials. The single-particle levels of the two dots can be modulated by time-dependently driven gate voltages. The on-site Coulomb interaction on each dot is assumed to be infinitely strong, while the intra-dot Coulomb interaction is finite (it determines the distance between the stable regions $(0,0)$ and $(1,1)$ in the stability diagrams in (b).) (b) Stability diagram for the weakly coupled double dot as function of the gate voltage applied to the two dots. Two possible driving cycles leading to the transfer of single electrons are indicated by blue lines. Generally, all gate voltage cycles  enclosing a triple point lead to quantized charge in the slow driving regime and can even result in plateaux in the heat transport. (c) Analogy of the driving cycle with the thermodynamic cycle of a Carnot engine. }
%\end{center}
\end{figure}

Here, we explain the working principle of one simple example system, where the analogue of a Carnot engine can be implemented using time-dependently driven quantum dots~\cite{Juergens2013Jun}. The sketch in Fig.~\ref{fig:cyclic}(a) shows the set-up with two quantum dots \textit{tunnel-coupled} to electronic reservoirs {\it and} to each other (in contrast to the dots studied in the section about the quantum-dot analogue of a thermal couple presented above, which were \textit{purely capacitively} coupled to each other). These electronic reservoirs can possibly be at different temperatures and/or electrochemical potentials and act as particle and heat baths. Depending on the gate voltage applied to the two quantum dots, this subsystem --- representing the heat engine --- is found in different (stable) charge configurations, see Fig.~\ref{fig:cyclic}(b). Some of these charge configurations are indicated in the figure; note that the degeneracy of each of these states depends on the electronic \textit{spin}. For example, while the state $(0,0)$ is non-degenerate, the state $(0,1)$ is doubly degenerate, since the electron occupying the right dot can have either spin up or spin down. It therefore has a Shannon entropy given by $\ln2$. Only at the boundaries of the stable regions, the occupation of the dot can change, and only at the so-called triple points where three stable regions touch, electron transport through the double dot can occur in a close-to-equilibrium situation. Importantly - in the weak coupling regime - electron-tunnelling at the \textit{lines} between two stable configurations, leading to a change of the double-dot's charge state, always occurs with \textit{one} of the reservoirs only.

%%%%%%%%%%%%%%%%%%%%%%%%%%%%%%%%%%%%%%%%%%%%%%%%%%%%%%%%%%%%%%%%%%%%%%%%%%%%%%%%%%
%%%%%%%%%%%%%%%%%%%%%%%%%%%%%%%%%%%%%%%%%%%%%%%%%%%%%%%%%%%%%%%%%%%%%%%%%%%%%%%%%%

We now consider a cyclic modulation of the gate voltages, which is typically used for the double dot operation as a quantized charge pump~\cite{Pothier1992,Chorley2012Apr,Roche2013Mar}. Such a double-dot pump can transfer charges even against an applied bias voltage and it can hence be viewed as a kind of ``battery charger"; it's efficiency depends on leakage currents and on the amount of heating occurring during the cycle. Interestingly, this pump can also be operated as a refrigerator or as a heat engine doing work on the AC fields, see Ref.~\cite{Juergens2013Jun} for details on the example discussed here. We now briefly explain the operational cycle of these machines, where we assume the driving to be infinitely slow. If the gate voltages, $V_\text{g,L}(t)$ and $V_\text{g,R}(t)$, are driven in time in such a way that one of the triple points is enclosed by the driving cycle --- see the example of the cycle around the upper triple point --- exactly one electron is pumped through the device per period. Interestingly, this goes along with the transfer of a \textit{quantized} amount of energy as well.  This result can be understood following the processes occurring along the driving cycle. Let us assume that we start the driving cycle in the stable region $(0,0)$. When crossing the line to the $(1,0)$ region, one electron is transferred from the left reservoir to the double dot, while also increasing its entropy from 0 to $\ln2$. Even though the total system might be in nonequilibrium (either due to a temperature or a potential gradient or both) nonetheless the Clausius relation applies here: this is due to the fact that coupling occurs to a \textit{single} reservoir at a time, leading to the occurrence of  an \textit{isothermal} process! The heat transferred between the reservoir and the double dot is hence  given by  $k_\text{B}T_\text{L}\ln 2$, with $T_\text{L}$ being the temperature of the left reservoir. The following part of the cycle where an electron is transferred from the left to the right dot takes place while the double dot is effectively decoupled from both reservoirs. This part of the cycle is hence \textit{adiabatic} in the thermodynamic sense. Finally a second isothermal transition occurs when the tunnelling process from  $(0,1)$ to $(0,0)$ occurs due to coupling to the right reservoir only, leading to the transferred heat $k_\text{B}T_\text{R}\ln 2$.
This sequence of processes leads to the transfer of quantized charge and heat and constitutes a close analogy to the Carnot cycle. 

Note that the transfer of heat proportional to $\ln2$ directly links to the well-known Landauer principle, stating that the erasure of a bit of information (here implemented in the electron spin) goes along with the energy cost of $k_\text{B}T\ln 2$. The consequences of this principle are further treated in Sec. V of this book. Since the transfer of heat is directly linked to the spin degeneracy of the quantum dot spin, switching on a magnetic field would lead to the complete suppression of the effect. 

In this infinitely slow driving regime --- neglecting electronic leakage (heat or charge) currents which are strongly suppressed here
--- a heat engine extracting heat from the hot reservoir and doing work on the AC fields, can be operated in the Carnot limit (and similarly a refrigerator would have an optimal coefficient of performance). Note however, that a nonvanishing driving frequency drastically reduces the efficiency~\cite{Juergens2013Jun}.  In contrast, the "battery charger" could operate at efficiencies of about $70\%$ of the optimal value even when the driving frequency is increased.

%\bigskip

%%%%%%%%%%%%%%%%%%%%%%%%%%%%%%%%%%%%%%%%%%%%%%%%%%%%%%%%%%%%%%%%%%%%%%%%%%%%%%%%%%
\section{Future directions}
%%%%%%%%%%%%%%%%%%%%%%%%%%%%%%%%%%%%%%%%%%%%%%%%%%%%%%%%%%%%%%%%%%%%%%%%%%%%%%%%%%

This chapter shows a few examples of thermoelectric effects and heat engines proposed and/or realized in different types of nanoelectronic systems. However, the thermodynamics of nanoelectronic devices is an active research field and numerous open questions are currently being addressed, or will be of interest in the future. 

On one hand, there is the ongoing quest for improving efficiencies of nano-thermoelectrics, which to date are very small, with the exception of a very recent experiment~\cite{Josefsson2018Jul}. The aim of such research is to improve device designs by increasing the desired electronic response and at the same time decreasing leakage (for example by introducing long photonic cavities~\cite{Bergenfeldt2014Feb} or using suspended structures~\cite{Bourgeois2005Feb,Zink2010Mar,Poran2017Mar,Cui2017Dec}).

On the other hand there are many fundamental questions (that could on the long term also lead to improved applications). For example, the nanostructures presented in this chapter involved either strong electron-electron interactions (Coloumb blockade) with weak coupling to the reservoirs,  or negligibly small electron-electron interactions with arbitrarily strong coupling to the reservoirs. However, 
the regime of strong electron-electron interactions with strong coupling to the reservoirs is of  great  interest in quantum thermodynamics.  In order to go to higher power outputs one is forced to consider stronger coupling. But the strong-coupling regime of nanoelectronic devices presents a challenge for theorists, see e.g., Refs.~\cite{Zitko2013Oct,Kennes2013Jun,Costi2010Jun,Strasberg2016Jul,Whitney2016Nov} and chapter 22 in this book for recent developments. 

Furthermore, as we  mentioned previously, the potential of quantum interference and entanglement for new types of heat engine has been little exploited so far, and constitutes an intriguing direction to explore. Also, the possibility of engineering the state of the bath, using entanglement or correlations as a resource for engines, is currently investigated and addressed in other chapters of this book.

\vskip 5mm
\acknowledgements

{\bf ACKNOWLEDGEMENTS.}  We acknowledge the support of the COST Action MP1209 ``Thermodynamics in the quantum regime'' (2013-2017), which enabled us to meet regularly to learn about and discuss much of the physics presented in this chapter. RW acknowledges the financial support of the French National Research Agency's ``Investissement d'avenir'' program (ANR-15-IDEX-02) via the Universit\'e Grenoble Alpes QuEnG project.  RS is supported by the Spanish Ministerio de Econom\'ia y Competitividad via the Ram\'on y Cajal program RYC-2016-20778. JS acknowledges support from the Knut and Alice Wallenberg foundation and from the Swedish VR.

%\bibliography{cite}


%%
%%%%%%%%%%%%%%%%%%%%%%%%%%%%%%%%%%%%%%%%%%%%%
%%%%%%%%%%%%%%%%%%%%%%%%%%%%%%%%%%%%%%%%%%%%%
%%%%%%%%%%%%%%%%%%%%%%%%%%%%%%%%%%%%%%%%%%%%%
%%%%%%%%%%%%%%%%%%%%%%%%%%%%%%%%%%%%%%%%%%%%%
%%
\end{document}